\documentclass[12pt]{article}

\usepackage{amssymb}
\usepackage{amsmath}
\usepackage{epsfig}

\renewcommand{\theequation}{\arabic{section}.\arabic{equation}}
\begin{document}

\title{\bf Isothermal Plasma Waves in Gravitomagnetic Planar Analogue}

\author{M. Sharif \thanks{e-mail: msharif@math.pu.edu.pk} and Umber
Sheikh
\\Department of Mathematics, University of the Punjab,\\Quaid-e-Azam
Campus Lahore-54590, Pakistan.}

\date{}

\maketitle

\begin{abstract}
We investigate the wave properties of the Kerr black hole with
isothermal plasma using 3+1 ADM formalism. The corresponding Fourier
analyzed perturbed GRMHD equations are used to obtain the dispersion
relations. These relations lead to the real values of the components
of wave vector $\textbf{k}$ which are used to evaluate the
quantities like phase and group velocities etc. These have been
discussed graphically in the neighborhood of the pair production
region. The results obtained verify the conclusion of Mackay et al.
according to which rotation of a black hole is required for negative
phase velocity propagation.
\end{abstract}


\section{Introduction}

The study of gravitomagnetic waves in rotating black hole is
important because the existence of black holes can be ultimately
verified with the help of infalling plasma radiation and
super-radiance of gravitomagnetic waves. Information from the
magnetosphere can be transmitted from one region to another only by
means of propagation allowed by plasma state. The gravitational
field and the wind bring perturbations to the external fluid
dynamics. The response of black holes to the external perturbations
can be explored with the help of wave scattering method.

General Relativity is the theory of four-dimensional spacetime but
we experience a three-dimensional space evolved in time. It is
easier to split the spacetime into three-dimensional space and
one-dimensional time to develop a better understanding of the
physical phenomenon. The split we usually use in understanding the
general relativistic physics of black holes and plasmas is 3+1 ADM
split introduced by Arnowitt et al. \cite{ADM}. This split in the
formalation of general relativity is particularly appropriate for
applications to the black hole theory as described by Thorne et al.
\cite{TM1}-\cite{TPM}. Using this formalism, the wave propagation
theory in the Friedmann universe was investigated by Holcomb and
Tajima \cite{HT}, Holcomb \cite{Ho} and Dettmann et al. \cite{De}.
Komissarov \cite{Kom} discussed the famous Blandford-Znajek
solution.

Blandford and Znajek \cite{BZ} found a process which describes the
extraction of rotational energy in the form of Poynting flux. The
black hole with a force free magnetosphere behaves as a battery with
internal resistivity in a circuit made by poloidal current. This
current system is considered to be equivalent to incoming and
outgoing waves. The incoming waves transport energy in a direction
opposite to the Poynting flux. Penrose \cite{P} was the pioneer who
gave the idea of extraction of energy from the rotating black hole
by a specific process called Penrose process. In the wavelength
analogue of Penrose process \cite{G} an incoming wave with positive
energy splits up into a transmitted wave with negative energy and a
refracted wave with enhanced positive energy. The negative energy
wave propagates into the black hole equivalent to a positive
Poynting flux coming out of the horizon \cite{Can}-\cite{Pre}.

The key features of the Kerr black hole were beautifully summarized
by M\"{u}ller \cite{Mu} who investigated the accretion physics in
the plasma regime of the general relativistic magnetohydrodynamics
(GRMHD). Punsley et al. \cite{Pu} considered the black hole
magnetohydrodynamics in a broader sense. Musil and Karas \cite{MK}
observed the evolution of disturbances originated in outer parts of
the accretion disk and developed a numerical scheme to show the
transmission and reflection of waves. Koide et al. \cite{Ko} modeled
the GRMHD behavior of plasma flowing into rotating black hole in a
magnetic field. They showed (numerical simulations) that energy of
the spinning black hole can be extracted magnetically. Zhang
\cite{Z1}-\cite{Z2} formulated the black hole theory for stationary
symmetric GRMHD with its applications in Kerr geometry. He discussed
wave modes for cold plasma with specific interface conditions. Buzzi
et al. \cite{BH1}-\cite{BH2} provided a linearized treatment of
transverse and longitudinal plasma waves in general relativistic two
component plasma (3+1 ADM formalism) propagating in radial direction
close to the Schwarzschild horizon.

Mackey et al. \cite{M} gave the idea that negative phase velocity
plane wave propagates in the ergosphere of a rotating black hole.
They verify that the rotation of a black hole is required for
negative phase velocity propagation which is a characteristic of
Veselago medium. This medium was hypothetically mentioned by
Veselago \cite{V} and later formed experimentally \cite{SSS} as a
material (called metamaterial or left-handed material). Much work
has been carried out to investigate the characteristics of this
medium \cite{many}. Woodley and Mojahedi \cite{WM} showed (using
full wave simulations and analytical techniques) that in
left-handed materials, the group velocity can be either positive
(backwards wave propagation) or negative. Sharif and Umber
\cite{S1}-\cite{S2} investigated some properties of plasma waves
by investigating real wave numbers. The analysis has been done for
the cold as well as isothermal plasmas living in the neighborhood
of the event horizon by using Rindler approximaton of the
Schwarzschild spacetime. In a recent paper, the same authors
\cite{S6} have found some interesting properties of cold plasma
waves using perturbation wave analysis to the GRMHD equations in
the vicinity of the Kerr black hole. They have also discussed the
existence of Veselago medium near the pair production region.

This paper has been extended to investigate the wave properties for
the isothermal plasma. We have focussed this work to investigate the
following three main objectives:
\begin{enumerate}
\item The behavior of gravitomagnetic waves under the
influence of gravity and magnetospheric wind is analysed. This helps
us to detect the response of the black hole magnetospheric plasma
oscillations to gravitomagnetic perturbations near the pair
production region. The pair production region lies near the event
horizon of the black hole.
\item The existence of Veselago medium in the black hole
regime is checked.
\item The negative phase velocity propagation regions are
investigated and compare the results with those obtained by Mackay
et al. \cite{M}.
\end{enumerate}
To this end, we derive the GRMHD equations in 3+1 formalism using
the isothermal equation of state. The component form of the
equations for specific background assumptions is obtained by using
perturbations. We consider the perturbed quantities as plane
harmonic waves produced by gravity and wind due to black hole
rotation. The Fourier analysis method for waves is applied and
dispersion relations are derived. These relations lead to the
$x$-component of the wave vector from which the relevant quantities
are investigated to analyze the wave properties near the pair
production region.

The paper is organized as follows. The next section is oriented with
the description of the Kerr analogue spacetime and mathematical laws
in 3+1 formalism for this model. Section \textbf{3} is devoted to
the assumptions corresponding to the background flow. In section
\textbf{4}, the GRMHD equations alongwith their Fourier analyzed
perturbed form for the isothermal equation of state of plasma are
given. Section \textbf{5} provides the solutions of dispersion
relations. In the last section, we shall discuss the results.

\section{Mathematical Framework}

This section contains the line element for a general spacetime
model. The electrodynamics corresponding to Kerr planar analogue in
3+1 formalism is also considered.

\subsection{Description of Model Spacetime}

The line element of the spacetime in 3+1 formalism can be written
as
\begin{equation}\label{1}
ds^2=-\alpha^2dt^2+\gamma_{ij}(dx^i+\beta^idt)(dx^j+\beta^jdt),
\end{equation}
where lapse function ($\alpha$), shift vector ($\beta$) and
spatial metric ($\gamma$) are functions of time and space
coordinates.

We consider the planar analogue of Kerr spacetime \cite{Z2}, i.e.,
\begin{equation}\label{K}
ds^2=-dt^2+(dx+\beta(z)dt)^2+dy^2+dz^2.
\end{equation}
Here $z,~x,~y$ and $t$ correspond to Kerr's radial $r$, axial
$\phi$, poloidal $\theta$ and time $t$ coordinates. The Kerr metric
depends non-trivially on both $r$ and $\theta$, whereas this model
metric depends on $z$ only. The lapse function $\alpha$ is taken to
be unity to avoid the effects of horizon and redshift. The value of
the shift function $\beta$ (analogue to the Kerr-type
gravitomagnetic potential) decreases monotonically from $0$ ($z
\rightarrow \infty$) to some constant value ($z \rightarrow
-\infty$). We have assumed the direction of $\beta$ along $x$-axis.
This shift function derives an MHD wind which extracts translational
energy analogous to the rotational energy for the Kerr metric. The
shift vector in three dimensions will be denoted by the Greek letter
$\beta.$ The Kerr-type horizon has been pushed off to $z=-\infty$.
The pair production region lies at $z=0$ where the plasma is
created. The newly created particles are then driven up to
relativistic velocities by magnetic-gravitomagnetic coupling as they
flow off to infinity and down towards the horizon. Geometrized units
will be used throughout the paper.

\subsection{Electrodynamics in Kerr Planar Analogue}

We consider the magnetosphere filled with MHD fluid and take the
perfect MHD flow condition in fluid's rest-frame
\begin{equation}
\textbf{E}+\textbf{V}\times\textbf{B}=0
\end{equation}
with $\textbf{V}$, $\textbf{B}$ and $\textbf{E}$ are fiducial
observer (FIDO) measured fluid velocity, magnetic and electric
fields respectively. For perfect MHD flow in (\ref{1}) with
$\alpha=1$, differential form of Faraday's law in 3+1 formalism
\cite{Z1} turn out to be
\begin{equation}\label{a}
\frac{d\textbf{B}}{d\tau}+(\textbf{B}.\nabla)\beta
-(\nabla.\beta)\textbf{B}=\nabla\times(\textbf{V}\times\textbf{B}),
\end{equation}
where $\frac{d}{d \tau}\equiv \frac{\partial}{\partial
t}-\beta.\nabla$ is the FIDO measured rate of change of any
three-dimensional vector in absolute space. Gauss law of magnetism
according to FIDO can be written as \cite{Z1}
\begin{equation}\label{b}
\nabla.\textbf{B}=0.
\end{equation}
For (\ref{1}) with $\alpha=1$, the local conservation law of
rest-mass \cite{Z1} according to FIDO is
\begin{equation}\label{c}
\frac{D\rho_0}{D\tau}+\rho_0\gamma^2\textbf{V}.\frac{D\textbf{V}}{D\tau}+
\rho_0\nabla.(\textbf{V}-\beta)=0,
\end{equation}
where $\rho_0$ is the rest-mass density, $\gamma$ is the Lorentz
factor and $\frac{D}{D\tau}\equiv\frac{d}{d\tau}+\textbf{V}.\nabla
=\frac{\partial}{\partial t}+(\textbf{V}-\beta).\nabla$ is the time
derivative moving along the fluid. The FIDO measured law of force
balance equation \cite{Z1} for the spacetime, given by Eq.(\ref{1})
with $\alpha=1$, takes the form
\begin{eqnarray}\label{d}
&&\left\{\left(\rho_0\gamma^2\mu+\frac{\textbf{B}^2}{4\pi}\right)\gamma_{ij}
+\rho_0\gamma^4\mu
V_iV_j-\frac{1}{4\pi}B_iB_j\right\}\frac{DV^j}{D\tau}
+\rho_0\gamma^2V_i\frac{D\mu}{D\tau}\nonumber\\
&&-\left(\frac{\textbf{B}^2}{4\pi}\gamma_{ij}
-\frac{1}{4\pi}B_iB_j\right){V^j}_{,k}V^k
=-\rho_0\gamma^2\mu\beta_{j,i}V^j-p_{,i}\nonumber\\
&&+ \frac{1}{4\pi}(\textbf{V}\times \textbf{B})_i
\nabla.(\textbf{V}\times \textbf{B})
-\frac{1}{8\pi}(\textbf{B})^2_{,i}+\frac{1}{4\pi}B_{i,j}B^j\nonumber\\
&&-\frac{1}{4\pi}[\textbf{B}\times\{\textbf{V} \times (\nabla
\times (\textbf{V} \times \textbf{B})-(\textbf{B}.\nabla ) \beta)
+(\textbf{V} \times \textbf{B}). \nabla \beta \}]_i,
\end{eqnarray}
where $\mu$ is the specific enthalpy and $p$ is the pressure of the
fluid. The FIDO measured local energy conservation law (Eq.(2.4) of
\cite{S2}), for Eq.(\ref{1}) with $\alpha=1$, is given as follows
\begin{eqnarray}\label{e}
&&\rho_0\gamma^2\frac{D\mu}{D\tau}+\mu\gamma^2\frac{D\rho_0}{D\tau}
+2\rho_0\mu\gamma^4\textbf{V}.\frac{D\textbf{V}}{D\tau}-\frac{d
p}{d\tau}-\mu\rho_0\gamma^2\nabla.\beta\nonumber\\
&&+\rho_0\mu\gamma^2(\nabla.\textbf{V})
-\rho_0\mu\gamma^2\textbf{V}.(\textbf{V}.\nabla)\beta
+\frac{1}{4\pi}\{(\textbf{V}\times\textbf{B}).(\nabla
\times \textbf{B})\nonumber\\
&&+(\textbf{V}\times\textbf{B}).\frac{d}{d\tau}(\textbf{V}\times\textbf{B})
+(\textbf{V}\times\textbf{B}).\{(\textbf{V} \times
\textbf{B}).\nabla\}\beta \nonumber\\
&&-(\textbf{V}\times\textbf{B}).
(\textbf{V}\times\textbf{B})(\nabla.\beta)\}=0.
\end{eqnarray}
Eqs.(\ref{a})-(\ref{e}) give the perfect
GRMHD equations.

\section{Specialization of Background Flow for Model Spacetime}

In this section, we give the background flow and relative
assumptions which will be used to simplify the problem.

\subsection{Description of Flow Quantities}

The FIDO measured 4-velocity of fluid can be described by a spatial
vector field lying in the $xz$-plane \cite{Z2}
\begin{equation*}
\textbf{V}=V(z)\textbf{e}_\textbf{x}+u(z)\textbf{e}_\textbf{z}.
\end{equation*}
Here the Lorentz factor takes the form
$\gamma=\frac{1}{\sqrt{1-u^2-V^2}}$. The magnetic field measured by
FIDO is also assumed to be in the $xz$-direction
\begin{equation*}
\textbf{B}=B\{\lambda(z)\textbf{e}_\textbf{x}
+\textbf{e}_\textbf{z}\},
\end{equation*}
where $B$ is constant. The corresponding Poynting vector becomes
\begin{equation*}\label{PV}
\textbf{S}=\frac{1}{4\pi}(\textbf{E}\times\textbf{B}).
\end{equation*}
We have considered an example of stationary flow of an isothermal
MHD fluid in our model spacetime (\ref{K}). These flows are used as
stationary model magnetospheres whose dynamical perturbations are to
be studied. The plasma is moving in the $xz$-direction. The
perturbed flow is along $z$-direction due to the black hole's
gravity and along $x$-direction due to rotation of the black hole
(in direction of shift vector of our analogue spacetime). This flow
will be analyzed to seek the properties of plasma waves.

\subsection{Perturbations and Wave Propagation}

The perturbed flow in the magnetosphere (which is in the $xz$-plane)
can be characterized by velocity \textbf{V}, magnetic field
\textbf{B}, the fluid density $\rho$ and pressure $p$. We denote the
unperturbed quantities by a superscript zero and the following
dimensionless notations are used for perturbations ($\delta
\textbf{V},~\delta \textbf{B},~\delta \rho,~\delta p$)
\begin{eqnarray}{\setcounter{equation}{1}}
\label{p} \tilde{\rho}&\equiv&\frac{\delta
\rho}{\rho}=\tilde{\rho}(t,x,z),\quad
\tilde{p}\equiv\frac{\delta p}{p}=\tilde{p}(t,x,z),\nonumber\\
\textbf{v}&\equiv& \delta
\textbf{V}=v_x(t,x,z)\textbf{e}_\textbf{x}+v
_z(t,x,z)\textbf{e}_\textbf{z},\nonumber\\
\textbf{b}&\equiv& \frac{ \delta
\textbf{B}}{B}=b_x(t,x,z)\textbf{e}_\textbf{x}
+b_z(t,x,z)\textbf{e}_\textbf{z}.
\end{eqnarray}
The perturbed variables take the following form
\begin{eqnarray}
\label{pv}
\rho=\rho^0+\rho\tilde{\rho},&~&p=p^0+p\tilde{p},\nonumber\\
\textbf{V}=\textbf{V}^0+\textbf{v},&~&
\textbf{B}=\textbf{B}^0+B\textbf{b}.
\end{eqnarray}
It is also assumed that the perturbations have sinusoidal
dependence of $t,~x$ and $z$. Thus
\begin{eqnarray}
\label{ps}
\tilde{\rho}(t,x,z)=c_1e^{-i(\omega
t-k_xx-k_zz)},&~&\tilde{p}(t,x,z)=c_6e^{-i(\omega
t-k_xx-k_zz)},\nonumber\\
v_x(t,x,z)=c_2e^{-i(\omega t-k_xx-k_zz)},&~&
v_z(t,x,z)=c_3e^{-i(\omega t-k_xx-k_zz)},\nonumber\\
b_x(t,x,z)=c_4e^{-i(\omega t-k_xx-k_zz)},&~&
b_z(t,x,z)=c_5e^{-i(\omega t-k_xx-k_zz)}.
\end{eqnarray}
Using the values of components of $\textbf{k}$, we can discuss the
quantities like phase velocity vector, group velocity vector,
refractive index and its change with respect to angular frequency.
These quantities would help to investigate the wave behavior of
the Kerr black hole magnetosphere and the properties of Veselago
medium.

\section{GRMHD Equations for Kerr Spacetime in Isothermal State of
Plasma}

The isothermal equation of state means that there is no exchange
of energy between the plasma and the magnetic field. This state
can be expressed by the following equation
\begin{equation*}
\mu=\frac{\rho+p}{\rho_0}=constant.
\end{equation*}
When we use this equation of state, the set of GRMHD
Eqs.(\ref{a})-(\ref{e}) take the following form for the spacetime
given in Eq.(\ref{K}), i.e, $\beta=(\beta_x,0,0)$
\begin{eqnarray}{\setcounter{equation}{1}}
\label{a1}
&&\frac{d\textbf{B}}{d\tau}+(\textbf{B}.\nabla)\beta
=\nabla\times(\textbf{V}\times\textbf{B}),\\
\label{b1}
&&\nabla.\textbf{B}=0,\\
\label{c1}
&&\frac{D(\rho+p)}{D\tau}+(\rho+p)\gamma^2\textbf{V}.\frac{D\textbf{V}}{D\tau}+
(\rho+p)\nabla.\textbf{V}=0,\\
\label{d1}
&&\left[\left\{(\rho+p)\gamma^2
+\frac{\textbf{B}^2}{4\pi}\right\}\delta_{ij}+(\rho+p)\gamma^4
V_iV_j-\frac{1}{4\pi}B_iB_j\right]\frac{dV^j}{d\tau}\nonumber\\
&&+(\rho+p)\gamma^2V_{i,k}V^k+(\rho+p)\gamma^4V_iV_j{V^j}_{,k}V^k
=(\rho+p)\gamma^2\beta_{j,i}V^j\nonumber\\
&&-p_{,i}+\frac{1}{4\pi}(B_{i,j}-B_{j,i})B^j
-\frac{1}{4\pi}\left\{\textbf{B}\times\left(\textbf{V}\times
\frac{d\textbf{B}}{d\tau}\right)\right\}_i,\\
\label{e1}
&&\gamma^2\textbf{V}.\frac{D}{D\tau}(\rho+p)
+2(\rho+p)\gamma^4\textbf{V}.\frac{D\textbf{V}}{D\tau}
-\frac{dp}{d\tau}+(\rho+p)\gamma^2(\nabla.\textbf{V})\nonumber\\
&&-(\rho+p)\gamma^2\textbf{V}.(\textbf{V}.\nabla)\beta
+\frac{1}{4\pi}\left[(\textbf{V}\times\textbf{B}).(\nabla
\times\textbf{B})\right.\nonumber\\
&&\left.+(\textbf{V}\times\textbf{B}).\frac{d}{d\tau}(\textbf{V}\times\textbf{B})
+(\textbf{V}\times\textbf{B}).\{(\textbf{V}\times\textbf{B}).\nabla\}\beta\right]=0.
\end{eqnarray}

These equations proceed in a similar way as used in
\cite{S1}-\cite{S2}. Equations (\ref{p}) and (\ref{pv}) as well as
the restrictions for the velocity and magnetic fields, given in
Section 3.1, lead to the perturbed form of Eqs.(\ref{a1})-(\ref{e1})
given in Appendix A. When we use Eq.(\ref{ps}), the Fourier analyzed
perturbed equations take the following form
\begin{eqnarray}\label{a4}
&&\iota k_z c_2-(\iota k_z\lambda+\lambda')c_3-c_4(\iota k_z
u-\iota\omega+u')\nonumber\\
&&+c_5\{(V-\beta)\iota k_z+(V-\beta)'\}=0,\\
\label{b4} &&\iota k_x c_2-\iota k_x\lambda c_3+\iota
c_5\{(V-\beta)k_x+k_zu-\omega\}=0,\\
\label{c4} &&k_xc_4=-k_zc_5,\\\label{d4}
&&c_1[\iota\rho\{-\omega+(V-\beta)k_x+uk_z\}-\{p'u+pu'+pu\gamma^2(VV'+uu')\}]
\nonumber\\
&&+c_2(\rho+p)\left[-\iota(\omega+\beta k_x)\gamma^2V+\iota
k_z\gamma^2uV +\iota k_x(1+\gamma^2V^2)\right.\nonumber\\
&&\left.+\gamma^2u\{(1+2\gamma^2V^2)V'+2\gamma^2uVu'\}\right]
+c_3(\rho+p)\left[-\iota(\omega+\beta k_x)\gamma^2u+\iota
k_x\gamma^2uV\right.\nonumber\\
&&\left.+\iota
k_z(1+\gamma^2u^2)-(1-2\gamma^2u^2)(1+\gamma^2u^2)\frac{u'}{u}
+2\gamma^4u^2VV'\right]+c_6[\iota p\{-\omega\nonumber\\
&&+(V-\beta)k_x+uk_z\}+\{p'u+pu'+pu\gamma^2(VV'+uu')\}]=0,\\
\label{e4} &&c_1\rho\gamma^2u\{(1+\gamma^2V^2)V'+\gamma^2uVu'\}
+c_6p[\gamma^2u\{(1+\gamma^2V^2)V'+\gamma^2uVu'\}+\iota
k_x]\nonumber\\
&&+c_2\left[-\iota(\omega+\beta
k_x)\left\{(\rho+p)\gamma^2(1+\gamma^2V^2)
+\frac{B^2}{4\pi}\right\}\right.\nonumber\\
&&\left.+\iota(k_xV+k_zu)\left\{(\rho+p)\gamma^2(1
+\gamma^2V^2)-\frac{B^2}{4\pi}\right\}
+(\rho+p)\gamma^4u\{(1+4\gamma^2V^2)uu'\right.\nonumber\\
&&\left.+4(1+\gamma^2V^2)VV'\}\right]
+c_3\left[-\iota(\omega+\beta k_x)\left\{(\rho+p)\gamma^4uV
-\frac{\lambda B^2}{4\pi}\right\}\right.\nonumber\\
&&\left.+\iota(k_xV+k_zu)\left\{(\rho+p)\gamma^4uV +\frac{\lambda
B^2}{4\pi}\right\}+(\rho+p)\gamma^2[2\gamma^2(1+2\gamma^2u^2)uVu'\right.\nonumber\\
&&\left.+\{(1+2\gamma^2u^2)(1+2\gamma^2V^2)
-\gamma^2V^2\}V']+\frac{B^2u\lambda'}{4\pi}\right]\nonumber\\
&&+\frac{B^2}{4\pi}c_4\{-\iota k_z(1-u^2)+uu'\}
+\frac{B^2}{4\pi}c_5\left\{-\lambda'-u(V-\beta)'+\iota k_x(1-V^2)\right.\nonumber\\
&&\left.-2\iota uVk_z\right\}=0,\\
\label{f4}
&&c_1\gamma^2\rho[u\{(1+\gamma^2u^2)u'+\gamma^2VuV'\}-V\beta']\nonumber\\
&&+c_6[\gamma^2p\{(1+\gamma^2u^2)uu'+
\gamma^2Vu^2V'-V\beta'\}+p'+p\iota k_z]\nonumber\\
&&+c_2\left[-\iota(\omega+\beta
k_x)\left\{(\rho+p)\gamma^4uV-\frac{\lambda
B^2}{4\pi}\right\}+\iota(k_xV+k_zu)
\left\{(\rho+p)\gamma^4uV\right.\right.\nonumber\\
&&\left.\left.+\frac{\lambda
B^2}{4\pi}\right\}+(\rho+p)\gamma^2\{\gamma^2u^2V'(1+4\gamma^2V^2)
-\beta'(1+2\gamma^2V^2)\right.\nonumber
\end{eqnarray}
\begin{eqnarray}
&&\left.+2V\gamma^2uu'(1+2\gamma^2u^2)\}\right]+c_3\left[-\iota(\omega+\beta
k_x)\left\{(\rho+p)\gamma^2(1+\gamma^2u^2)\right.\right.\nonumber\\
&&\left.\left.+\frac{\lambda^2B^2}{4\pi}\right\}
+\iota(k_xV+k_zu)\left\{(\rho+p)\gamma^2(1+\gamma^2u^2)
-\frac{\lambda^2B^2}{4\pi}\right\}\right.\nonumber\\
&&\left.+(\rho+p)\gamma^2[u'(1+\gamma^2 u^2)(1+4\gamma^2
u^2)+2u\gamma^2\{(1+2\gamma^2u^2)VV'-V\beta'\}]\right.\nonumber\\
&&\left.-\frac{B^2}{4\pi}\lambda\lambda'
u\right]+\frac{B^2}{4\pi}c_4\left\{\iota k_z\lambda
(1-u^2)+\lambda'-\lambda uu'\right\}+\frac{B^2}{4\pi}c_5\{2\lambda
uV\iota k_z\nonumber\\
&&+\lambda u(V-\beta)'-\lambda \iota k_x(1-V^2)]=0,\\\label{g4}
&&c_1\gamma^2[-\iota\omega\rho+\rho'u+\rho u'+2\rho u
\gamma^2(VV'+uu')-\rho \gamma^2uV\beta'+\iota
k_x\rho(V-\beta)\nonumber\\
&&+\rho\iota k_zu]+c_6[-\iota\omega
p(\gamma^2-1)+\gamma^2\{p'u+pu'+2pu\gamma^2(VV'+uu')\nonumber\\
&&-p\gamma^2uV\beta'+\iota k_xp(V-\beta)+p\iota
k_zu\}]+c_2[-\iota\omega\{2(\rho+p)\gamma^4V\nonumber\\
&&-\frac{B^2}{4\pi}(u\lambda-V)\}+\iota
k_x[(\rho+p)\gamma^2\{1+2\gamma^2V(V-\beta)\}
-\frac{B^2}{4\pi}(V-\beta)(u\lambda-V)]\nonumber\\
&&+\iota k_zu\{2(\rho+p)\gamma^4V+\frac{B^2}{4\pi}(u\lambda-V)\}
+(\rho+p)\gamma^2u\{2\gamma^2V'\nonumber\\
&&+6\gamma^4V(VV'+uu')-\beta'(1+2\gamma^2V^2)\}
-\frac{B^2\lambda'}{4\pi}]+c_3[-\iota\omega\{2(\rho+p)\gamma^4u\nonumber\\
&&+\frac{\lambda B^2}{4\pi}(u\lambda-V)\}+\iota
k_x(V-\beta)\{2(\rho+p)\gamma^4u+\frac{B^2\lambda}{4\pi}(u\lambda-V)\}\nonumber\\
&&+\iota k_z\{(\rho+p)\gamma^2(1+2\gamma^2u^2)-\frac{B^2\lambda
u}{4\pi}(u\lambda-V)\}
+(\rho+p)\gamma^2\{-\frac{u'}{u}\nonumber\\
&&+2\gamma^2uu'+6\gamma^4u^2(VV'+uu')+\gamma^2(VV'+uu')
-V\beta'(1+\gamma^2u^2)\}\nonumber\\
&&+\frac{B^2\lambda'}{4\pi}\{\lambda-u(u\lambda-V)\}]
+c_4\frac{B^2}{4\pi}[u\{\lambda'-(u\lambda-V)u'\}\nonumber\\
&&+\iota k_z(u\lambda-V)(1-u^2)]+c_5\frac{B^2}{4\pi}
\{-\lambda'V+u(u\lambda-V)(V-\beta)'\nonumber\\
&&-\iota k_x(u\lambda-V)(1-V^2)+2\iota k_z uV\}=0.
\end{eqnarray}
Eq.(\ref{c4}) gives the relation between $x$ and $z$-components of
the wave vector i.e., $k_z=-\frac{c_4}{c_5}k_x$ which will be used
in the next section.

\section{Numerical Solutions}

This section is devoted to the numerical solutions of the
dispersion equations. The following subsection contains the
relative assumptions which make the equations easier to deal with.

\subsection{Relative Assumptions}

In our stationary symmetric background, the $x$-component of the
velocity vector can be written in the form \cite{Z2} $V=C+\lambda
u,$ where $C\equiv\beta+V_F$ with $V_F$ as an integration constant.
We assume the value of the shift function \cite{Z2}
$\beta=\tanh(z)-1$ with $V_F=1$. Further, $B^2=8\pi$ and $\lambda=1$
are taken so that the magnetic field becomes constant. Thus the
$x$-component takes the form
$$V=1+\beta+u=\tanh(z)+u.$$

Substituting the value of $V$ in the conservation law of rest-mass
for three-dimensional hypersurface, i.e.,
\begin{equation}{\setcounter{equation}{1}}\label{clm}
\rho_0 \gamma u=\mu(\rho+p)\gamma u=A~(constant)
\end{equation}
with the assumption that rest-mass density is constant, we get an
equation of the form
$$3u^2+2u\tanh(z)+\tanh^2(z)-1=0$$
quadratic in $u$ with the assumption that $A/\rho_0=1$. The
solutions of this equation and the corresponding values of $V$ are
given as follows
\begin{eqnarray}\label{u1}
u_1&=&-\frac{1}{3}\tanh(z)-\frac{1}{3}\sqrt{3-2\tanh^2(z)},\nonumber\\
V_1&=&\frac{2}{3}\tanh(z)-\frac{1}{3}\sqrt{3-2\tanh^2(z)},\\
\label{u2}
u_2&=&-\frac{1}{3}\tanh(z)+\frac{1}{3}\sqrt{3-2\tanh^2(z)},\nonumber\\
V_2&=&\frac{2}{3}\tanh(z)+\frac{1}{3}\sqrt{3-2\tanh^2(z)}.
\end{eqnarray}
We shall use these values to solve the dispersion relations. The
Poynting vector for these values takes the following form
$$\textbf{S}=2\tanh(z)(\textbf{e}_\textbf{x}-\textbf{e}_\textbf{z}).$$
These quantities are valid for the region outside the pair
production region. We consider the region $-5\leq z\leq5$ and omit
the region $-1\leq z\leq1$ due to large variations in the background
flow quantities. In the rest of the region, these quantities become
constant and Fourier analyzed procedure is valid for this region.
Further, we use the relation $k_z=-k_x$ which reduces the wave
vector to $(k_x,0,-k_x)$.

The computer programming (using \emph{Mathematica}) is used to
evaluate a root of the dispersion relation for the plasma moving
towards the black hole with the velocity components given by
Eq.(\ref{u1}). This is given as a separate file with all the
required codes. Other roots can be obtained in a similar manner.

It is observed that the sextic equation has all roots admitting
imaginary values at several points. The quintic equation gives one
real root of the positive $z$ region for both the values of
velocity (Eqs.(\ref{u1}) and (\ref{u2})) shown in the Figures 1
and 2. For the negative $z$ region, the velocity, given by
Eq.(\ref{u1}), gives one real root shown in the Figure 3. The
approximated root becomes imaginary at $z=-5$ which we omit and
our mesh reduces to $-4.8\leq z\leq-1,~0\leq\omega\leq 10$ for the
interpolating function. The values at $z=-5$ are extrapolated
afterwards. The velocity components, given by Eq.(\ref{u2}) for
negative $z$ region, leads to three real roots shown in the
Figures 4, 5 and 6. The real data values for the root give a real
interpolation function.

It is clear that the Figures 1 and 2 represent the neighborhood of
the pair production region towards the outer end (as the wave number
is found for the positive values of $z$) whereas the Figures 3, 4, 5
and 6 show the neighborhood of the pair production region towards
the event horizon (as the wave number is determined for the negative
values of $z$).

\subsection{Results}

First, we obtain $k_x$ for the velocity components, given by
Eqs.(\ref{u1}) and (\ref{u2}) in the positive $z$ region. These are
shown in the Figures \textbf{1} and \textbf{2} respectively.

\begin{figure}
\center \epsfig{file=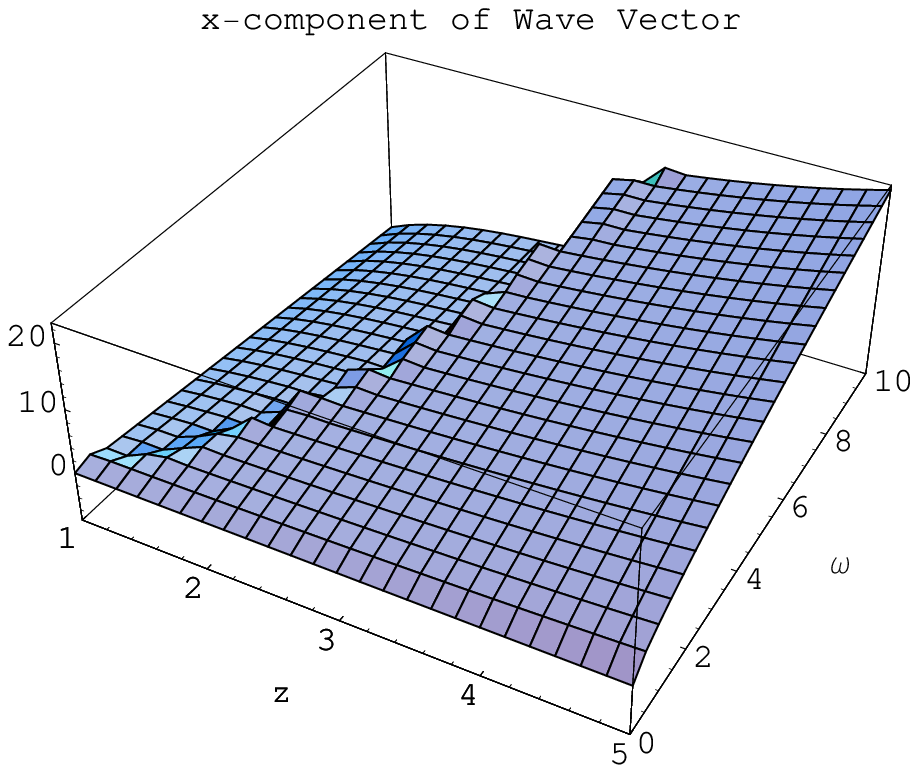,width=0.4\linewidth} \center
\epsfig{file=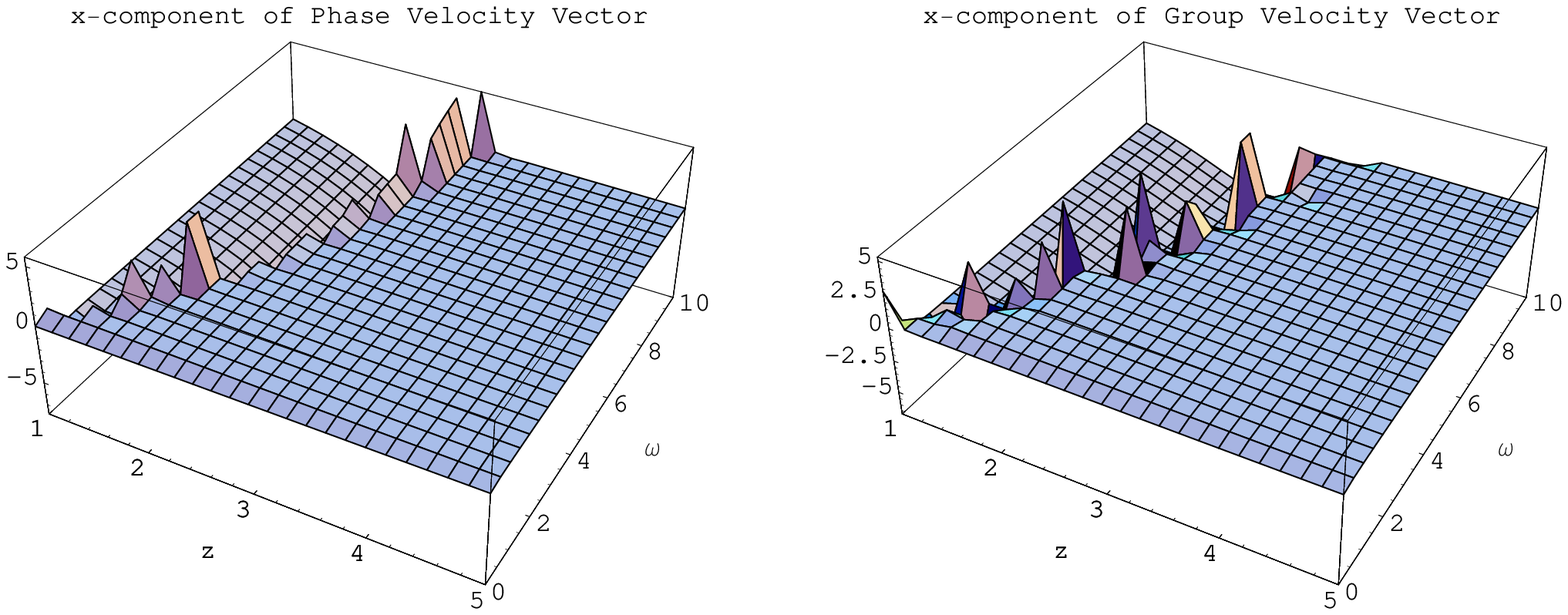,width=0.8\linewidth} \center
\epsfig{file=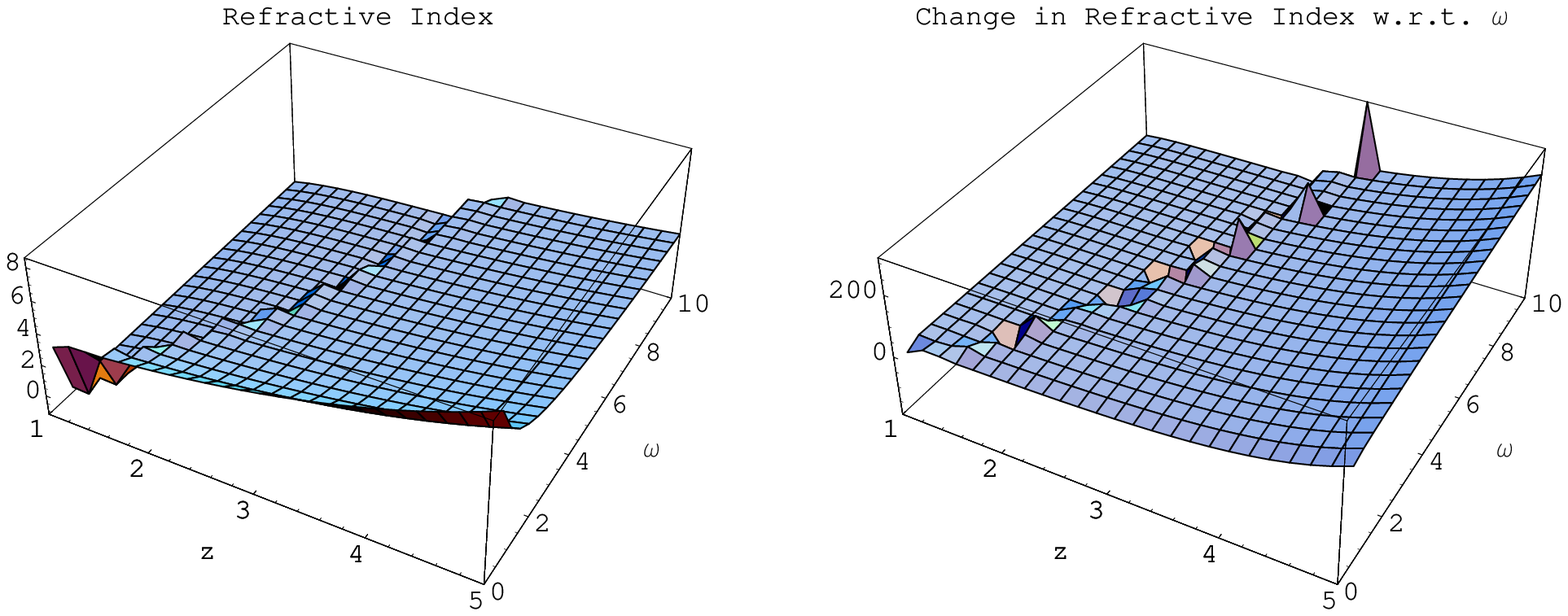,width=0.8\linewidth} \caption{For the velocity
components given by Eq.(\ref{u1}), plasma admits the properties of
Veselago medium near the pair production region. As the waves move
away from the pair production region, they disperse normally.
Negative phase and group velocity propagation regions are observed
near the pair production region.}
\end{figure}
In the Figure \textbf{1}, the $x$-component of the wave vector is
negative near the pair production region and for the waves with
negligible angular frequency. For each angular frequency, the waves
grow monotonically in number when they move away from the event
horizon. There is a sudden increase in the $x$-component of the wave
vector, it admits positive values for a particular value of $z$,
then decreases a bit and smoothly increases afterwards. The fluid
near the pair production region (region with negative $x$-component
of the wave vector) possesses the negative values for the
$x$-components of the phase and group velocities. For this region,
the wave vector is in the opposite direction of the Poynting vector
and hence it shows the existence of Veselago medium there \cite{V}.
For the same region, the phase and group velocity vectors are in the
direction opposite to the Poynting flux and hence the regions are of
negative phase and group velocity propagation. The change in the
refractive index with respect to the angular frequency is positive
for the regions (i) $1.6 \leq z<2,~0.4\leq\omega\leq1$, (ii) $2\leq
z<3,~0.42\leq\omega\leq2.92$, (iii) $3\leq
z<4,~0.225\leq\omega\leq10$ and (iv) $4\leq
z\leq5,~0.14\leq\omega\leq10$ for which the dispersion is normal
\cite{WM}, \cite{Ja}. In the rest of the region, most of the points
admit anomalous dispersion.

\begin{figure}
\center \epsfig{file=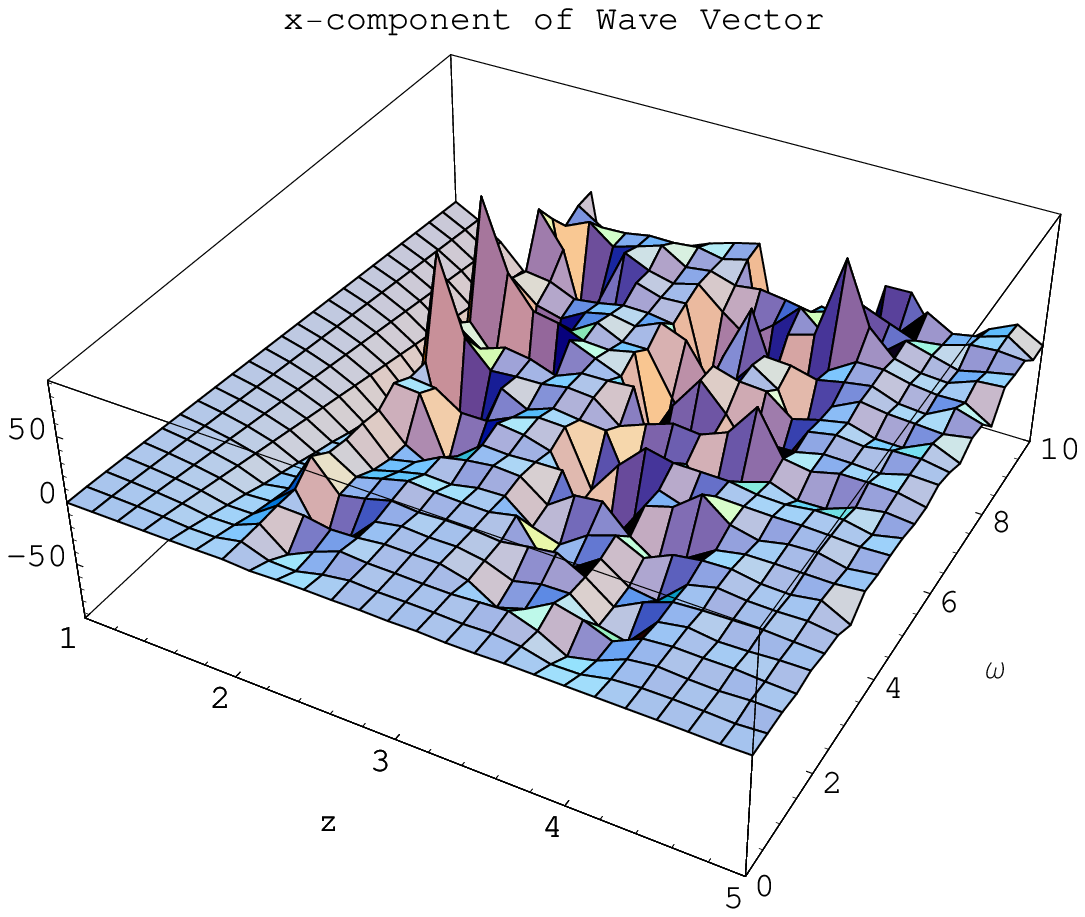,width=0.4\linewidth} \center
\epsfig{file=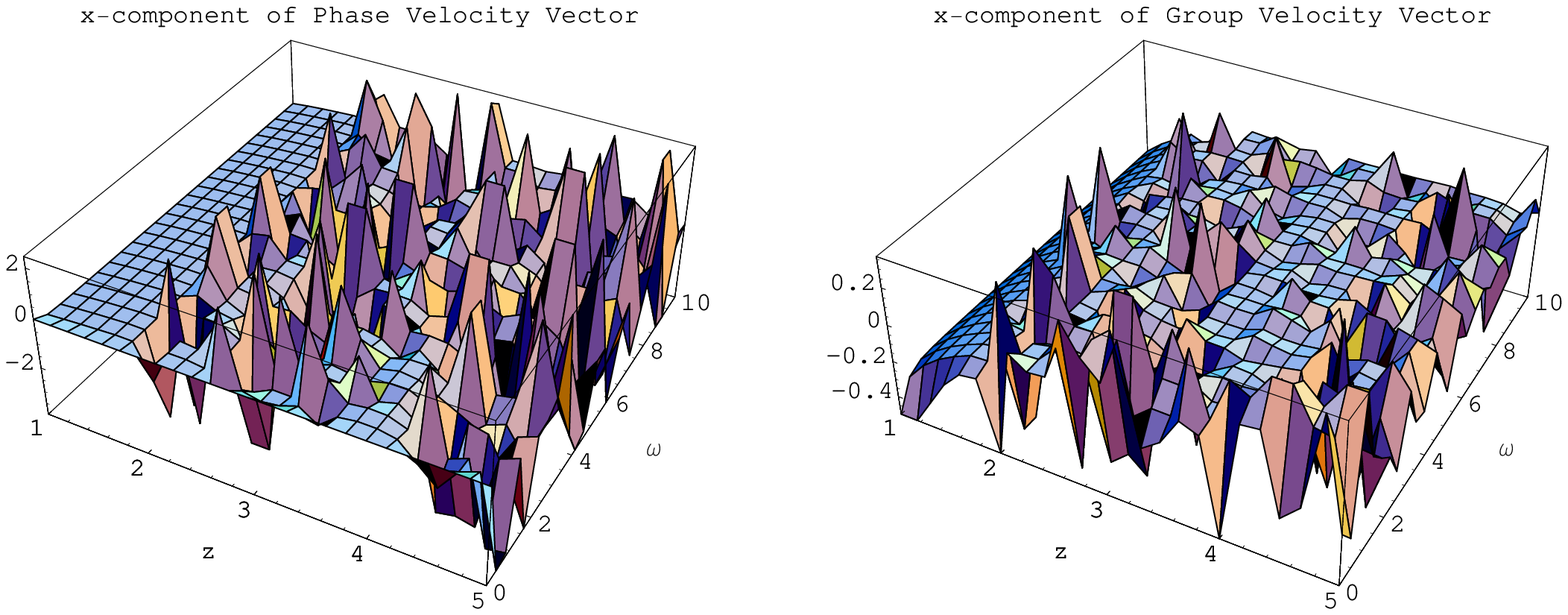,width=0.8\linewidth} \center
\epsfig{file=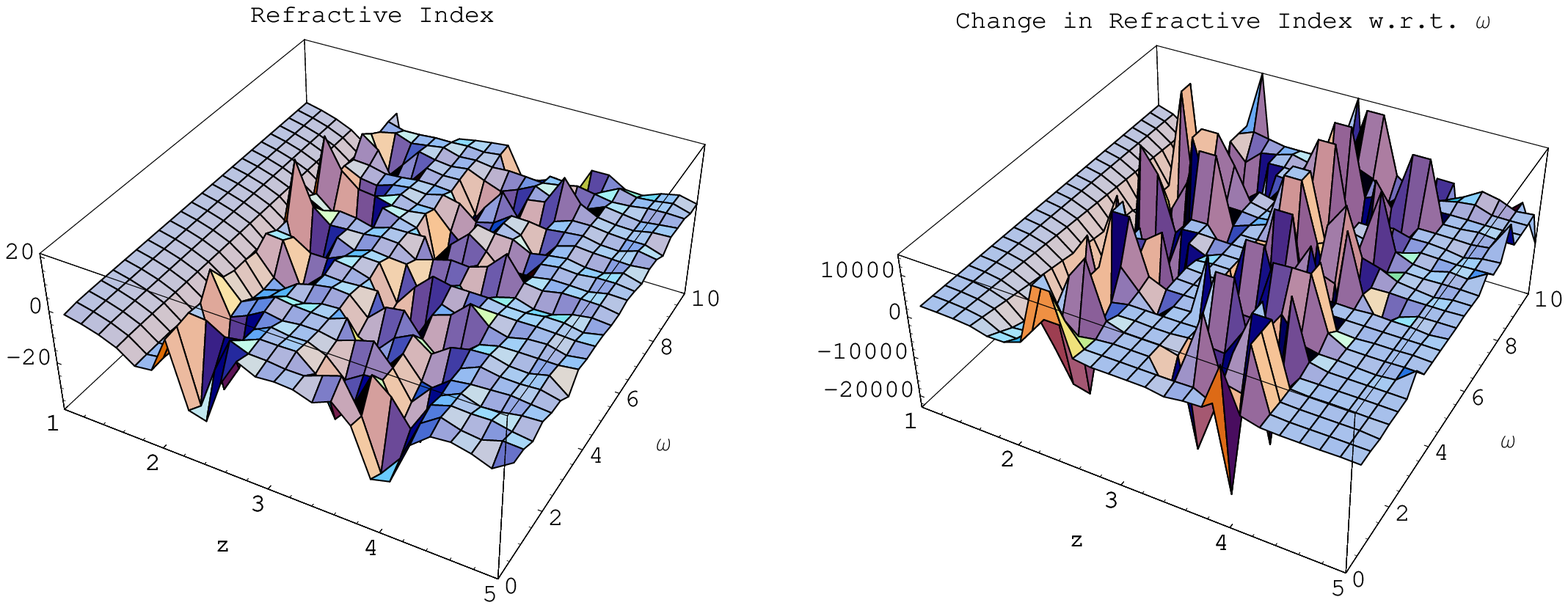,width=0.8\linewidth} \caption{Veselago medium
exists near the pair production region for the velocity components
given by Eq.(\ref{u2}). In the same region, negative phase and group
velocity propagation regions are observed. Most of the region in the
neighborhood of the pair production region shows anomalous
dispersion.}
\end{figure}
The Figure \textbf{2} shows that the $x$-component of the wave
vector is negative for the region $1.0\leq z\leq 1.89$. It is large
near the event horizon and decreases up to $z=1.75$ after which it
increases and fluctuations occur in the values. In the region
$1.89<z\leq10$, it takes random values. The negative values of $k_x$
in the region implies that the wave vector is in the opposite
direction to the Poynting vector which indicates the properties of
Veselago medium. In the region $0\leq z\leq1.89$, the $x$-components
of the phase and group velocities take negative values and hence
this is of negative phase and group velocity propagation region.
Both these quantities admit random values in the region $1.89\leq
z\leq10$. For the region $1\leq z\leq 1.6,~0<\omega\leq 0.079$, the
change in the refractive index with respect to the angular frequency
is positive and hence the dispersion is found to be normal. In the
region $1\leq z\leq1.6,~0.079<\omega\leq10$, the quantity
$\frac{dn}{d\omega}<0$ which indicates anomalous dispersion in this
region \cite{WM}. The rest of the region shows random points of
normal as well as anomalous dispersion.

For the negative $z$ region, i.e., the region towards the event
horizon of the black hole in the neighborhood of the pair production
region, we obtain one value of $k_x$ for the velocity components
given by Eq.(\ref{u1}) and three for the velocity components given
by Eq.(\ref{u2}). These values are shown respectively by the Figures
3, 4, 5 and 6.

\begin{figure}
\center \epsfig{file=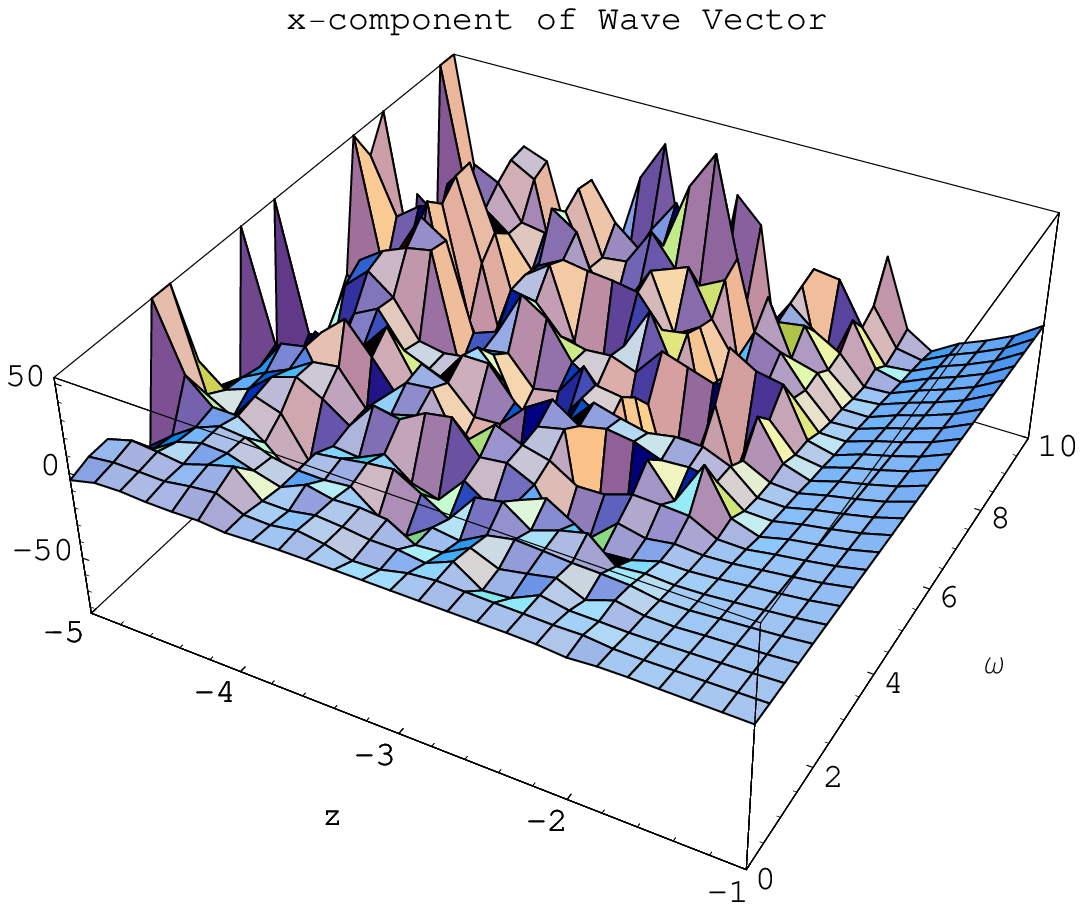,width=0.4\linewidth} \center
\epsfig{file=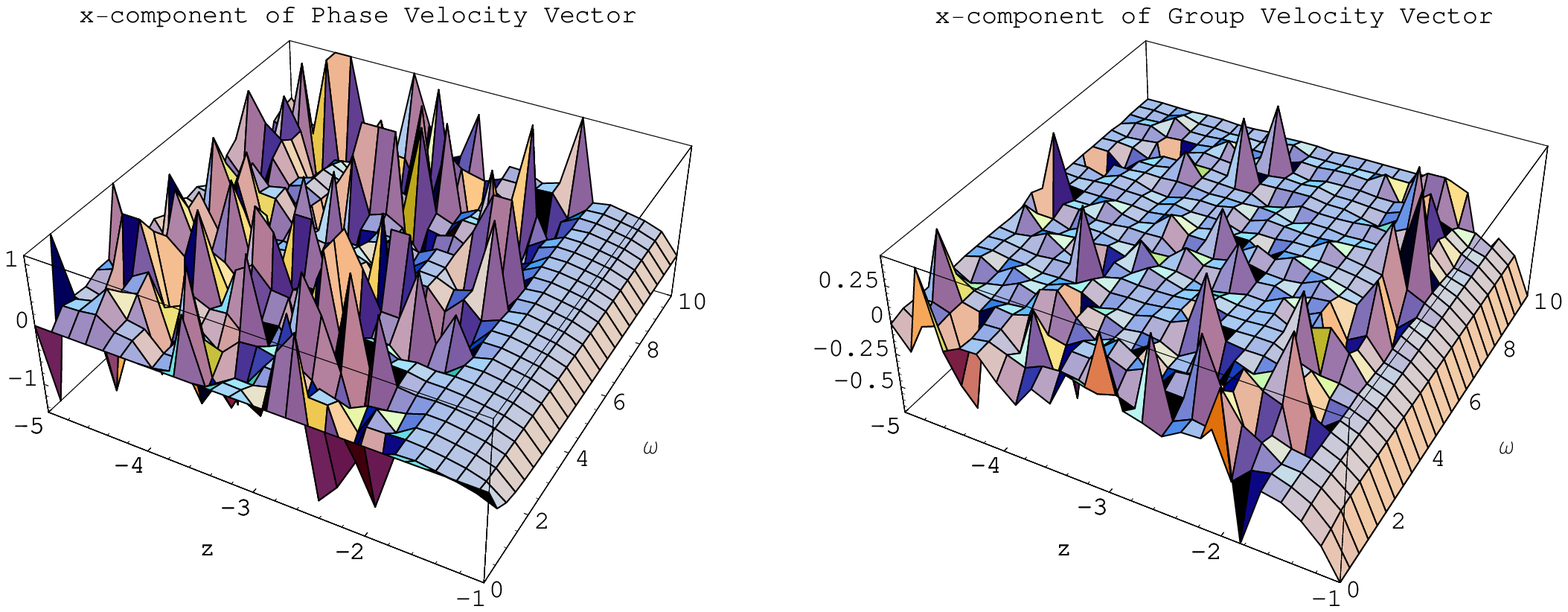,width=0.8\linewidth} \center
\epsfig{file=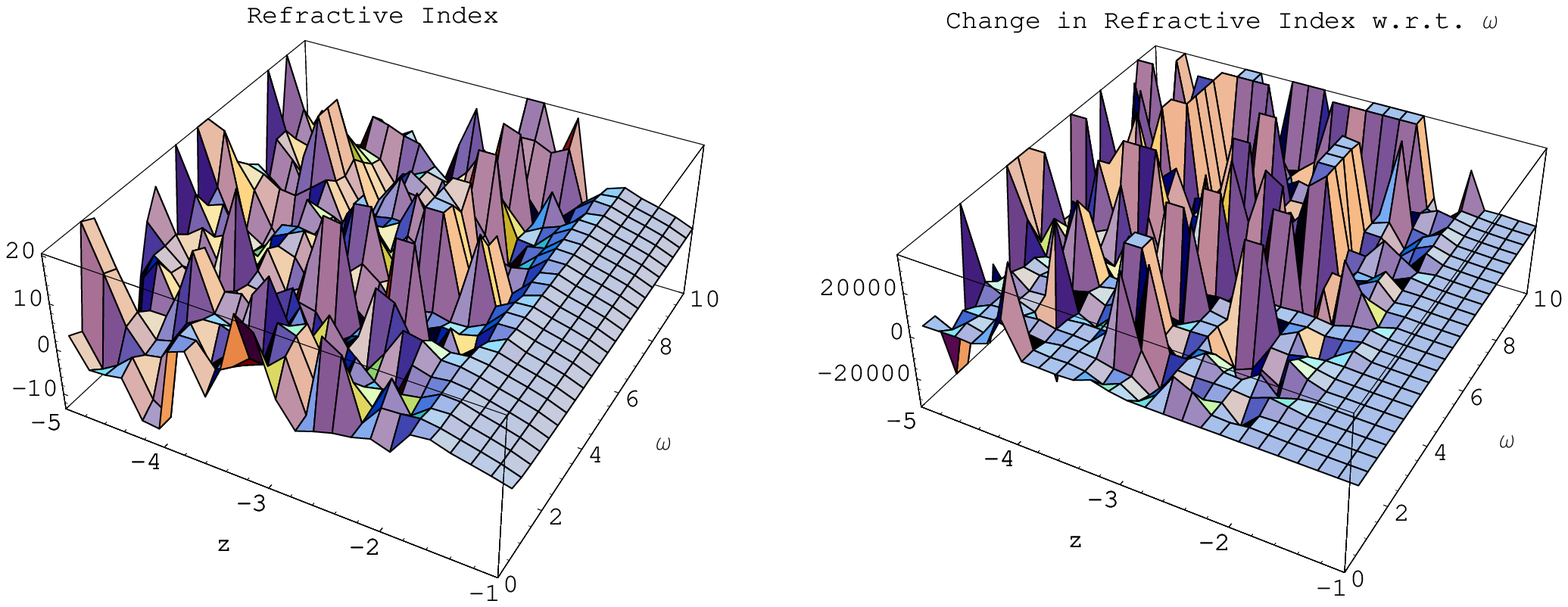,width=0.8\linewidth} \caption{Near the pair
production region, the dispersion is normal whereas the rest of the
region admits normal as well as anomalous points of dispersion for
velocity components given in Eq.(\ref{u1}).}
\end{figure}
In the Figure \textbf{3}, the $x$-component of the wave vector is
negative for the region $-1.925\leq z\leq -1.0$ where the Poynting
vector is parallel to the wave vector and hence the medium is usual.
The refractive index greater than one and positive change in the
refractive index with respect to the angular frequency indicate
normal dispersion. In the rest of the region, all the three
quantities admit random values and hence there are normal as well as
anomalous points of dispersion. For the region $-1.4\leq z\leq
-1.0,~0.5\leq\omega\leq10$, the $x$-components of phase and group
velocities are negative such that $v_{px}>v_{gx}$. These velocity
components admit random values in rest of the region.

\begin{figure}
\center \epsfig{file=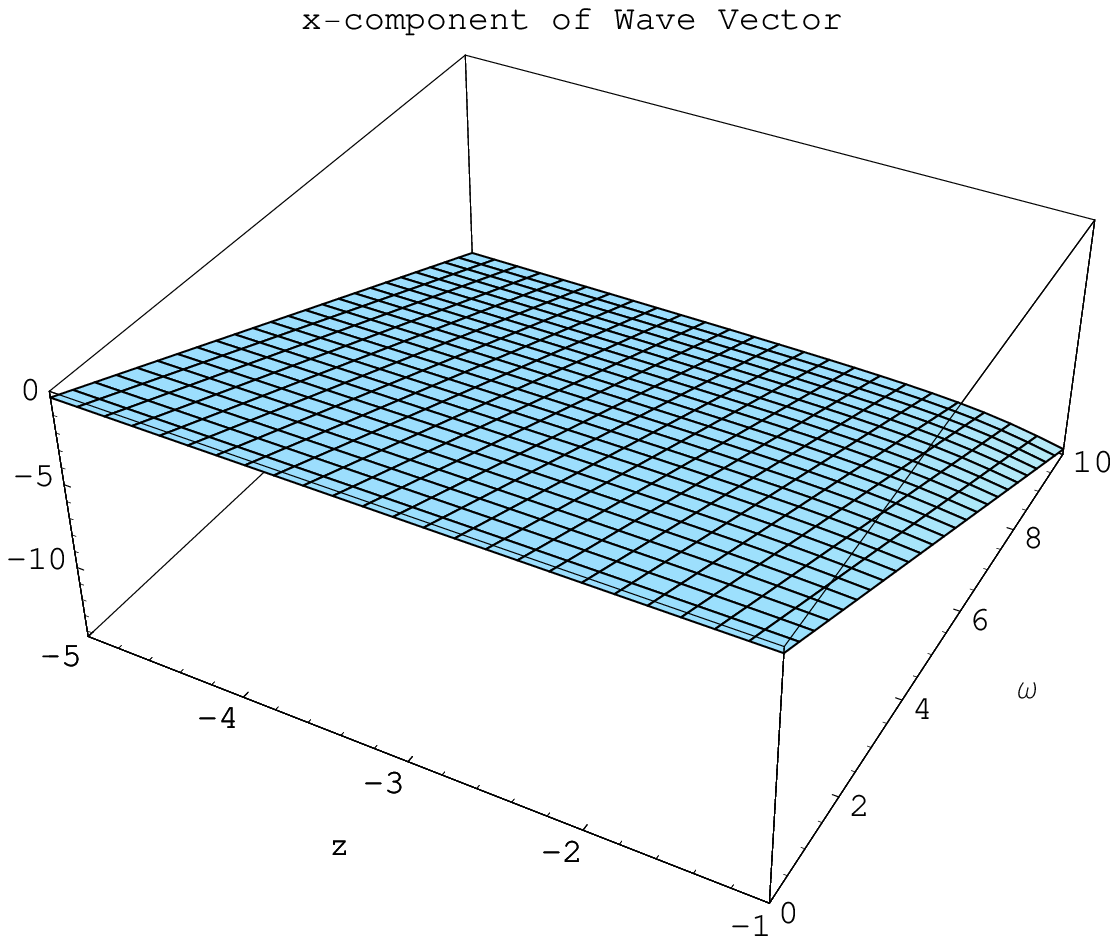,width=0.4\linewidth} \center
\epsfig{file=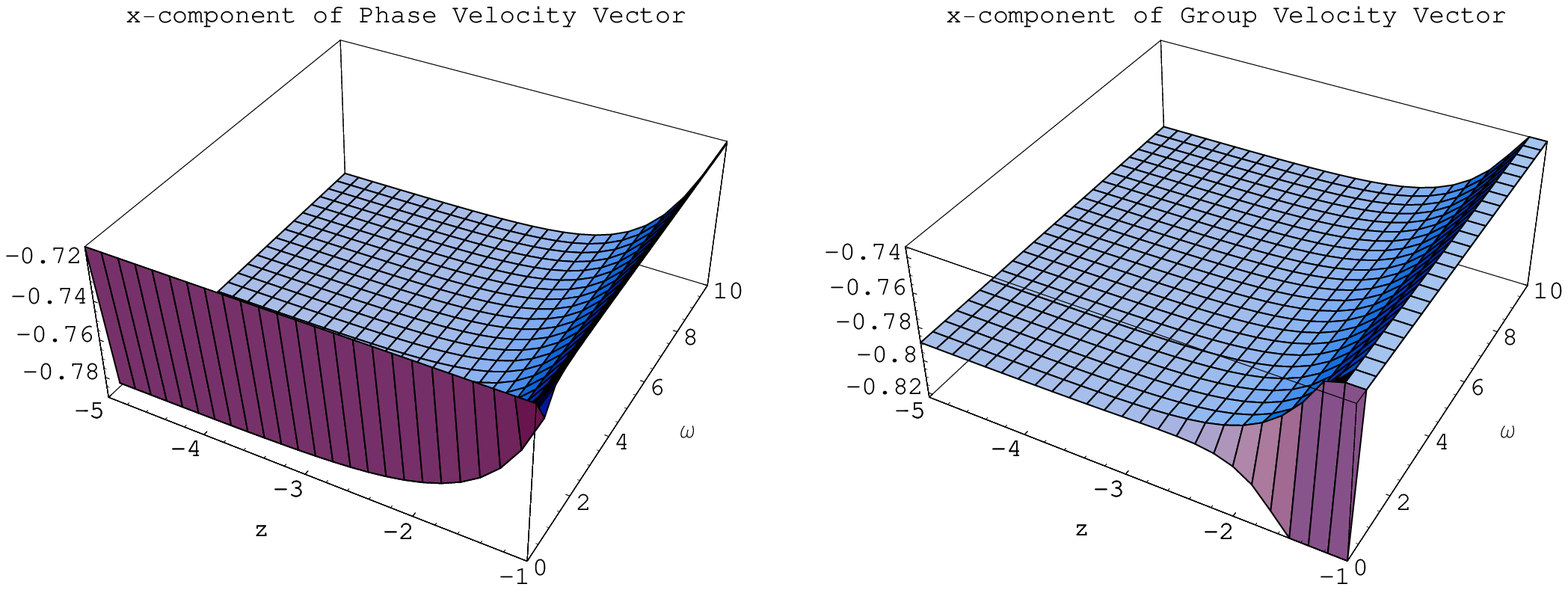,width=0.8\linewidth} \center
\epsfig{file=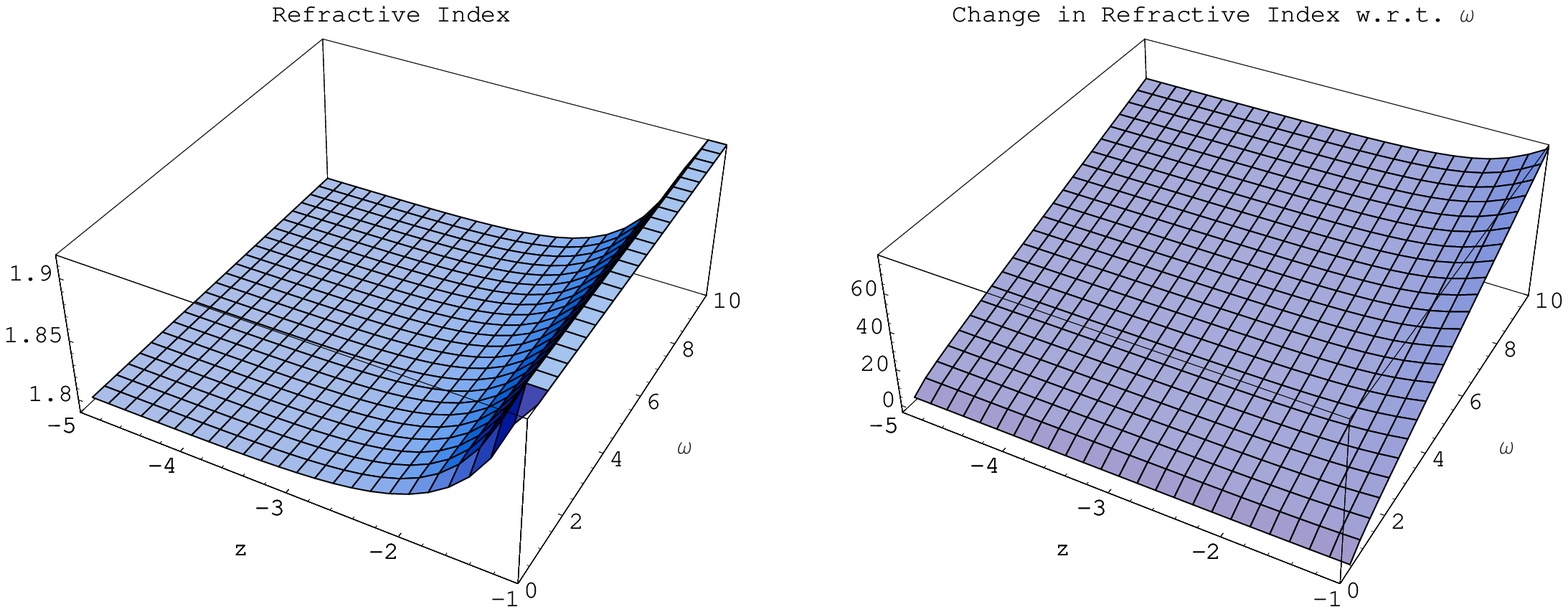,width=0.8\linewidth} \caption{For the velocity
components given by Eq.(\ref{u2}), dispersion is normal except for
the waves with very low angular frequency.}
\end{figure}
The Figure \textbf{4} indicates that the $x$-component of the wave
vector is negative for the whole region and hence the wave vector
and the Poynting vector are in the same direction showing the
existence of the usual medium. As the values of $z$ and $\omega$
grow, $k_x$ decreases and hence $v_{px}$ and $v_{gx}$ are negative
in this region. Although $v_{px}>v_{gx}$ for the region $0\leq
\omega\leq 10^{-15}$, yet they are nearly equal for rest of the
region. The refractive index is greater than one in the whole
region. The refraction increases as the waves move towards the pair
production region. The change in the refractive index with respect
to the angular frequency is negative for the region
$0<\omega\leq0.6$ which shows that anomalous dispersion of waves.
For the rest of the region, it is positive and thus indicates normal
dispersion.

\begin{figure}
\center \epsfig{file=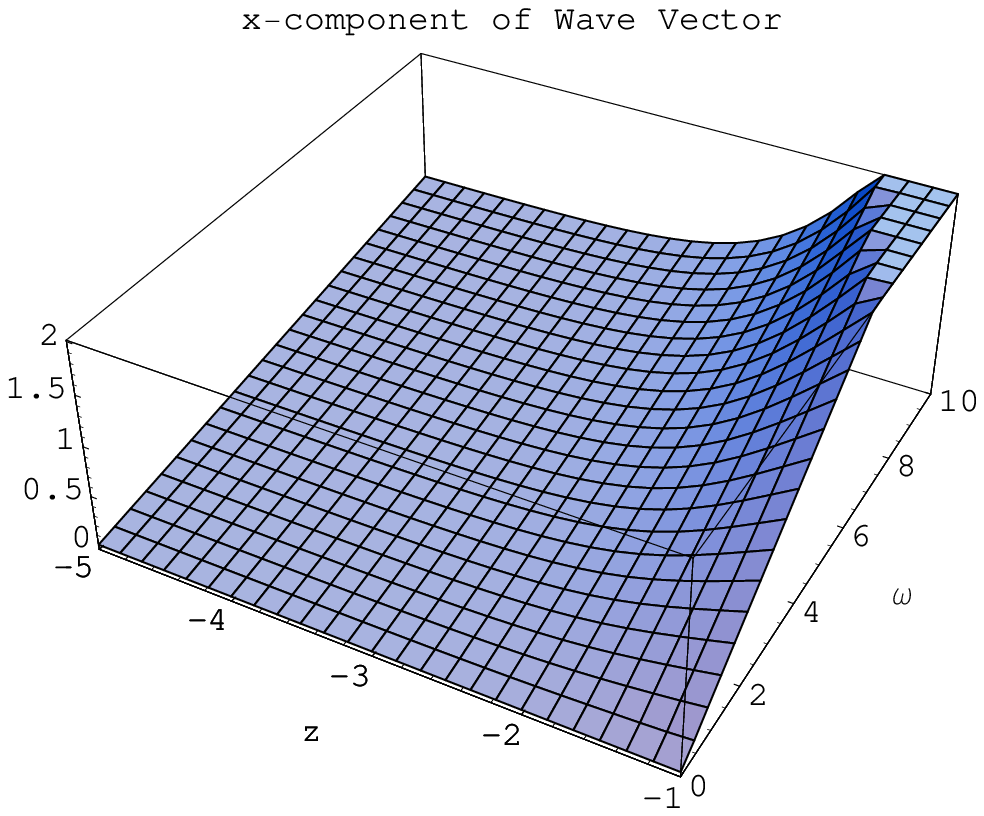,width=0.4\linewidth} \center
\epsfig{file=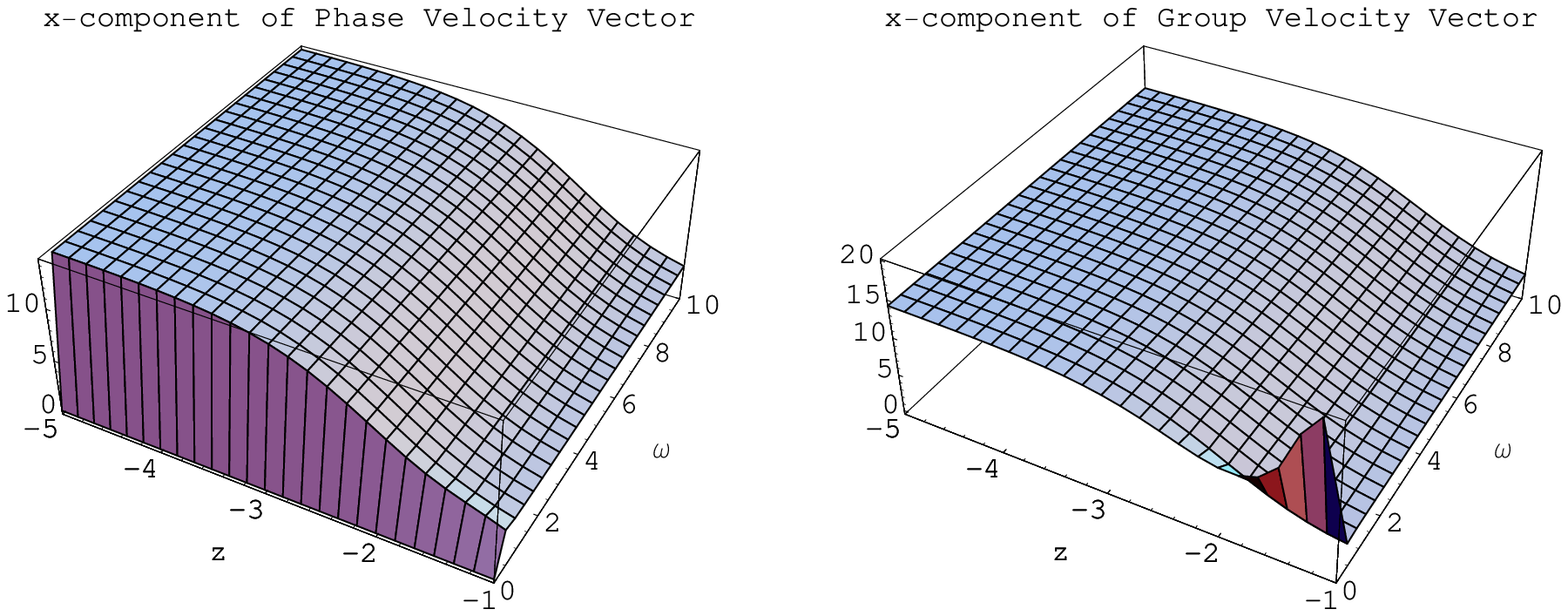,width=0.8\linewidth} \center
\epsfig{file=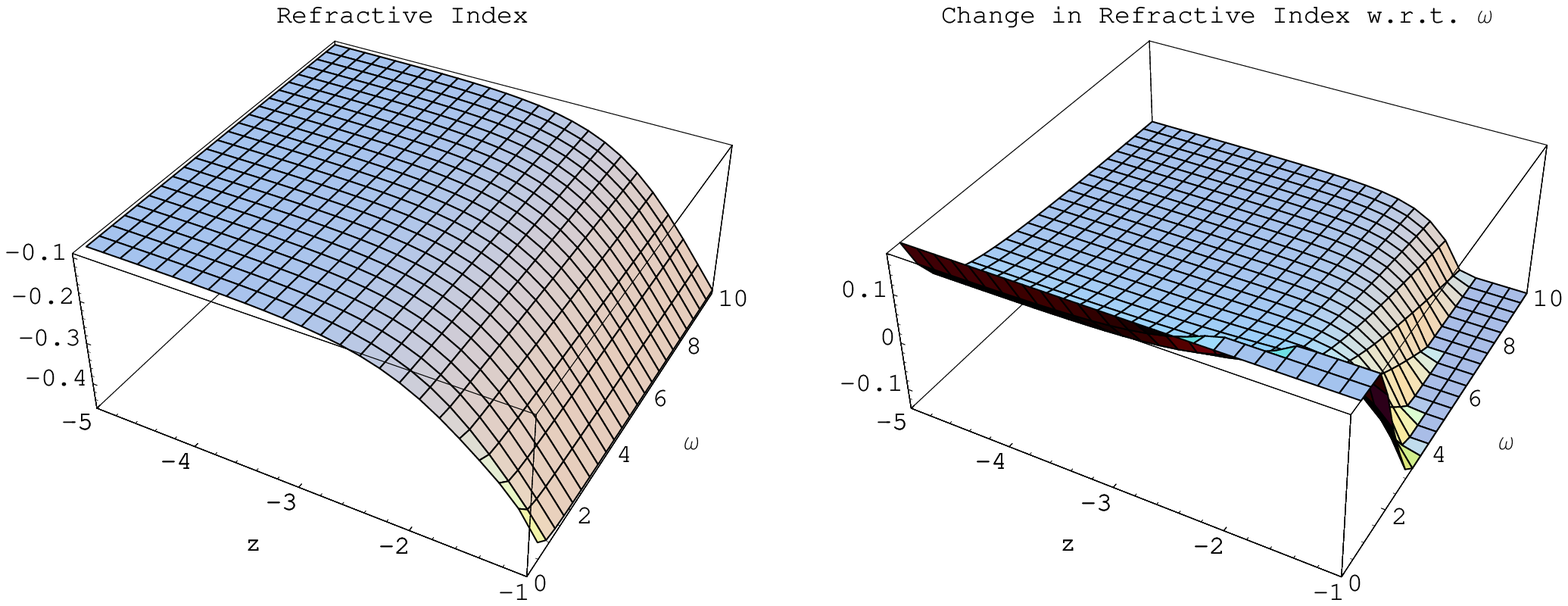,width=0.8\linewidth} \caption{For the velocity
components given by Eq.(\ref{u2}), Veselago medium exists in the
whole region with negative phase and group velocity propagation
property. The region of normal dispersion extends as the waves move
away from the pair production region. The waves with negligible
angular frequency do not disperse normally.}
\end{figure}
The Figure \textbf{5} shows that $k_x$ is positive for the whole
region. Thus the wave vector is in the opposite direction to the
Poynting vector which indicates the presence of Veselago medium. The
quantity $k_x$ increases with the increase in $z$ and $\omega$
except for the waves with negligible angular frequency. $v_{px}$ and
$v_{gx}$ are nearly equal and admit positive values which show
negative phase and group velocity propagation in the whole region.
The refractive index is negative and decreases as the values of $z$
and $\omega$ increase. The change in the refractive index with
respect to the angular frequency is negative for the regions (i)
$-2\leq z\leq-1,~2.15\leq\omega\leq10$ (ii) $-3\leq
z<-2,~5.5\leq\omega\leq10$ and (iii) $-4\leq
z<-3,~9.5\leq\omega\leq10$ which indicates anomalous dispersion in
these regions. It is positive for the rest of the region which
indicates normal dispersion except for the waves with negligible
angular frequency.

\begin{figure}
\center \epsfig{file=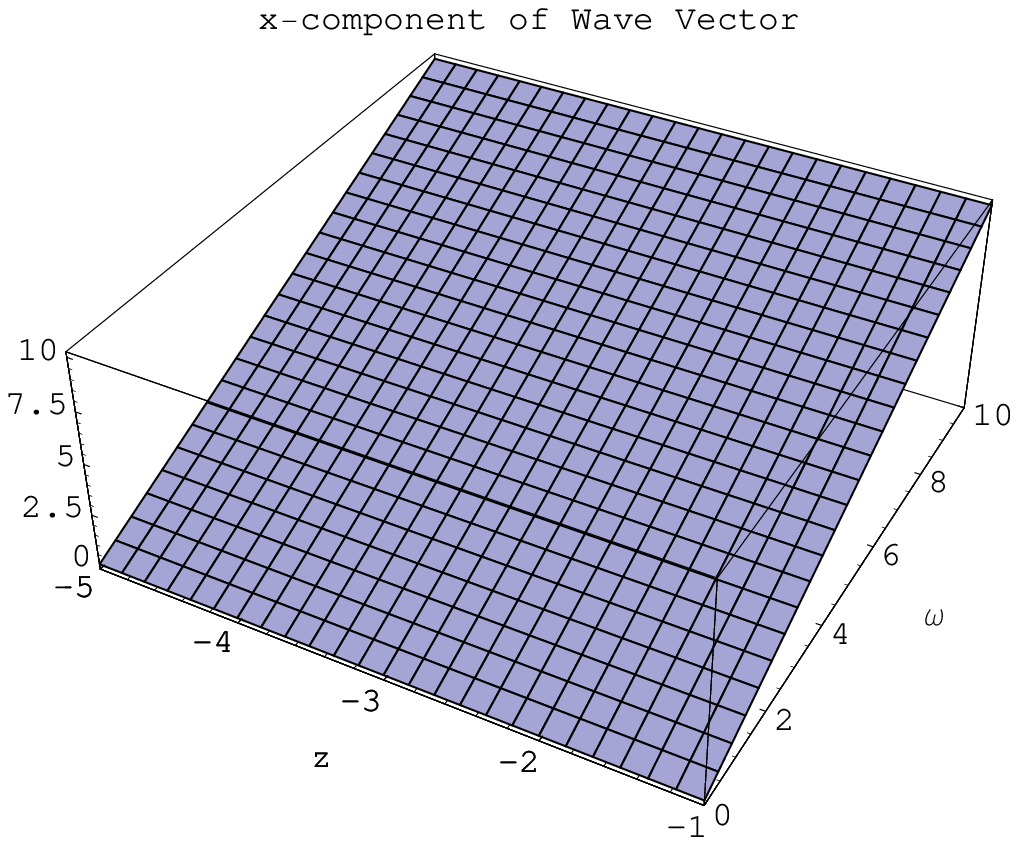,width=0.4\linewidth} \center
\epsfig{file=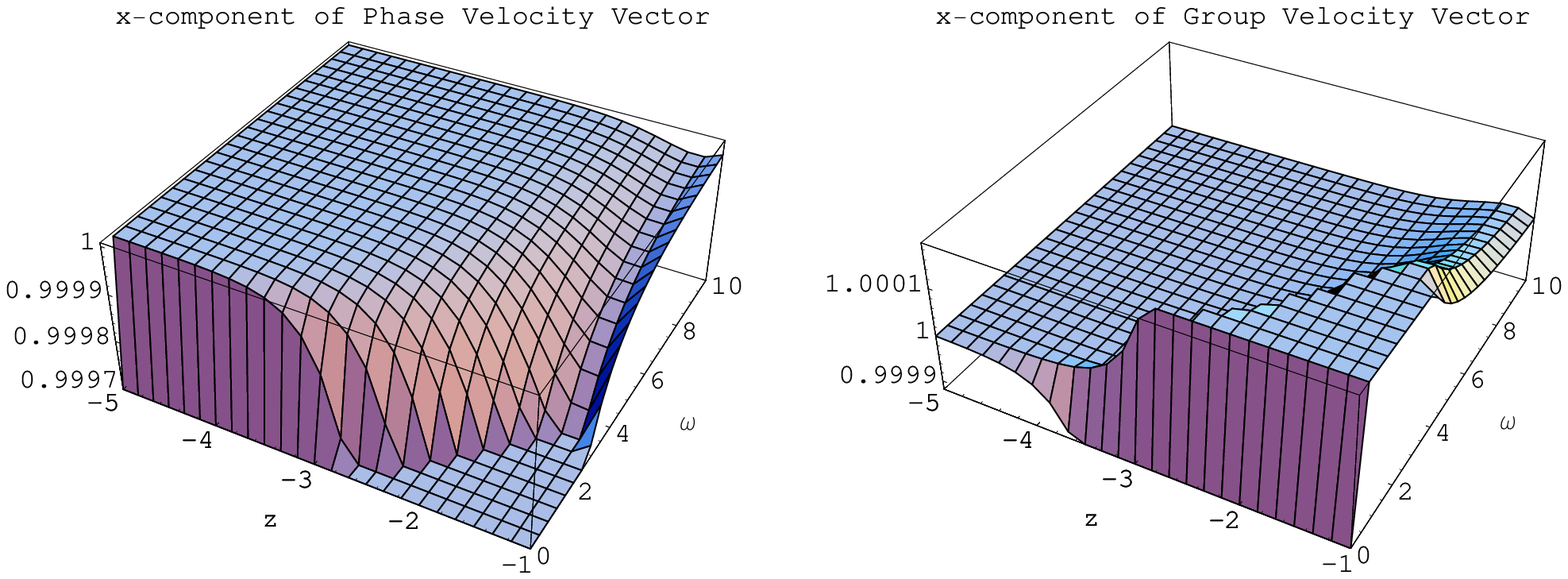,width=0.8\linewidth} \center
\epsfig{file=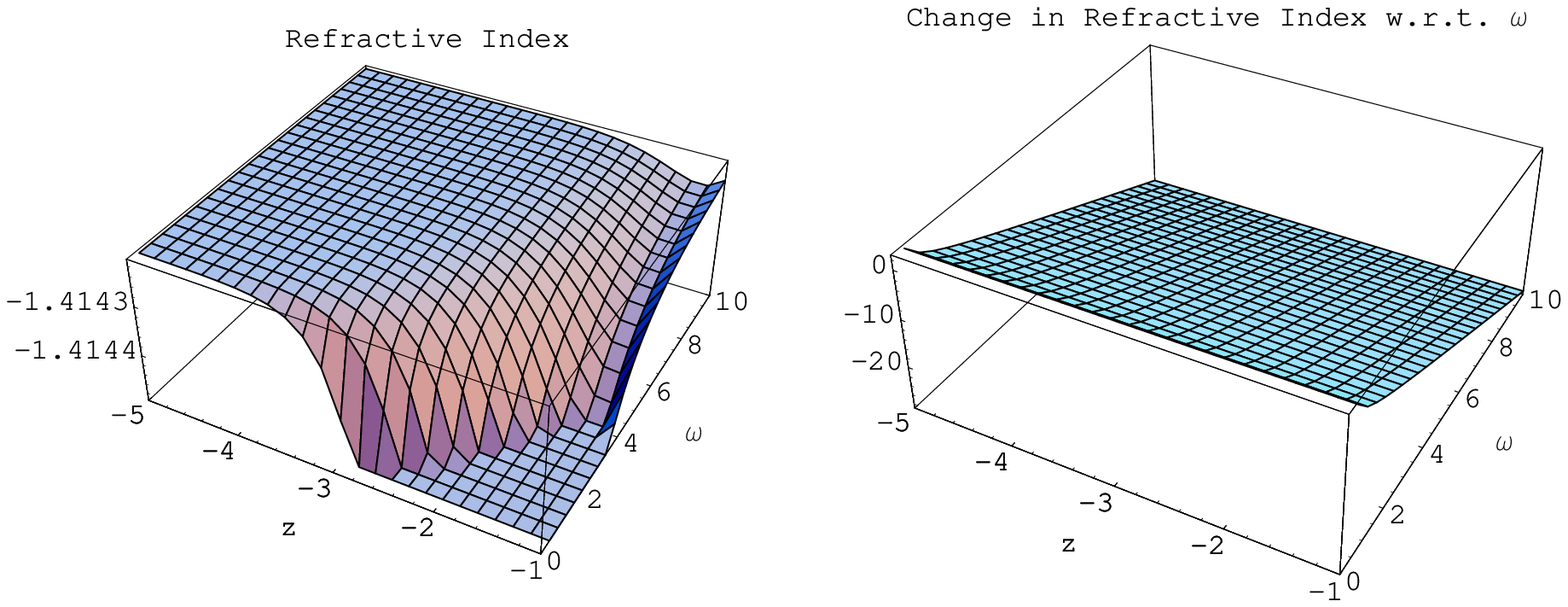,width=0.8\linewidth} \caption{For the velocity
components given by Eq.(\ref{u2}), plasma admits the properties of
Veselago medium. The dispersion is anomalous except for the waves
admitting negligible angular frequency. The negative phase and group
velocity propagation regions are also observed.}
\end{figure}
In the Figure \textbf{6}, the $x$-component of the wave vector is
positive for the whole region and increases with the increase in the
angular frequency. As the waves move away from the pair production
region, the $x$-component of the phase velocity decreases slightly
and then increases. In contrast, the $x$-component of the group
velocity increases a little and then decreases. The refractive index
is negative due to the fact that the Poynting vector is in the
opposite direction to the wave vector which shows the existence of
Veselago medium. The positive values of $x$-components of phase and
group velocities show the existence of negative phase and group
velocity propagation regions. The change in the refractive index
with respect to the angular frequency is negative throughout the
region and hence shows anomalous dispersion except for the waves
with negligible angular frequency.

\section{Conclusion}

It is well-known that charged particles are created in the pair
production region. Some of these particles which are pushed on to
orbits with negative energy by the Lorentz force move towards the
event horizon and others move towards the outer end of the
magnetosphere. These particles would reach their destinations if the
plasma existing in the neighborhood of the pair production region
allows them to do so. The generation of plasma is necessary to
support the MHD magnetically dominated flow. Due to particle
generation, waves are produced in the neighborhood of the pair
production region. The dispersion relations of waves lead to
understand how much the surrounding medium let the waves to disperse
through.

This paper studies the wave properties for the isothermal plasma
moving with velocity $\textbf{V}$ and admits a constant magnetic
field which thread the Kerr black hole magnetosphere. The
gravitomagnetic waves and the pair of particles are produced in the
$z=0$ region. If the medium living around the pair production region
allows the particles and waves to pass through, the energy
extraction from the black hole is possible. This can be well
understood by investigating properties of the waves in this region.

We have considered a black hole immersed in a rarefied plasma with
uniform magnetic field which seems to provide support for carrying
currents flowing across the magnetic field lines. Due to the
strength of the magnetic field, a lot of energy can be extracted due
to the plasma particles that fall into the black hole's horizon has
negative energy. The 3+1 GRMHD equations are derived for the
neighborhood of pair production region and two-dimensional
perturbations are discussed in the context of perfect MHD condition.
We assume that the rotation is in the $x$-direction and the horizon
is at $z=-\infty$. The perturbations are taken only in the
$xz$-direction. The dispersion relations are formulated by assuming
the perturbations as plane waves. We solve these relations by taking
the wave vector as $(k_x,0,-k_x)$ and obtain the $x$-component of
the wave vector. This component leads to properties of the
isothermal plasma in the neighborhood of the pair production region.

We have discussed these relations for the regions $1\leq z\leq5$ and
$-5\leq z\leq-1$. The dispersion relations for the region $1\leq
z\leq5$ are shown in the Figures 1 and 2. These Figures indicate
that near the pair production region, the plasma admits the
properties of Veselago medium. The region which is nearer to the
pair production region does not allow the waves to pass through.
Thus the particles and waves cannot get out of this region. The
small regions far away from the pair production region admit normal
dispersion of waves which indicate that the waves pass through them.
As we go away from the pair production region, normal dispersion
exists frequently as shown in the Figure 1.

The region $-5\leq z\leq-1$ allows us to investigate whether there
is a possibility for the waves to move towards the black hole event
horizon or not. The dispersion relations for this region are shown
in the Figures 3, 4, 5 and 6. From the Figures 3, 4 and 5, we find
that there are chances for the waves to pass through the
neighborhood of the pair production region when the plasma admits
the properties of usual or Veselago medium. The Figure 6 indicates
that there can be situation when plasma admits the properties of
Veselago medium in the neighborhood of the pair production region,
it may not allow the waves to pass through the region.

It is interesting to note that the Figures 2 and 3 show the
irregular dependence of wave vectors on the angular frequency and
$z$. Mathematically, this irregularity is due to the nature of the
roots obtained for these graphs. The irregular behavior may be due
to a disturbance of the equilibrium between outward and inward
directed forces. The outward directed forces are caused by the
particle pressure and the curvature drift due to non-uniform
magnetic field and inward directed forces are exerted by the
tangential stress of the magnetic field lines for the low frequency
regime.

For the high frequency regime, there is a class of MHD instabilities
which sometimes develop in a thin plasma column carrying a strong
current. If a kink begins to develop in such a column, the magnetic
forces on the inside of the kink become larger than those on the
outside, which leads to the growth of perturbation. The column then
becomes unstable and causes a disruption. Both the ballooning and
kink modes are ideal MHD instabilities.

In the Figures 1, 2, 5 and 6, we obtain the properties of Veselago
medium. The phase and group velocity vectors propagate in the
direction opposite to the Poynting vector which verify the results
of Mackay et al. \cite{M} according to which rotation of a black
hole is required for the negative phase velocity propagation.

We can conclude that waves produced in the pair production region
due to pair creation cannot get out of its neighborhood towards the
outer end of the magnetosphere. The same result has been obtained
for the cold plasma case \cite{S6}. We obtain some cases where
favorable conditions are present to allow energy to move towards the
black hole horizon. For the cold plasma, these conditions are
present for the usual medium whereas for the isothermal plasma,
these situations occur for usual as well as Veselago medium. For the
plasmas with pressure, the black hole can suck particles and waves
for both the usual and Veselago medium whereas for the cold plasmas,
this situation holds for the usual medium.

The strong magnetic coupling enforce the accreting particles to fall
into the black hole with negative energy and negative angular
momentum. This indicates that energy and angular momentum flow from
the black hole into the disk. When the particles fall into the black
hole with negative energy, energy is extracted from the black hole
\cite{AL}. When the particle with positive energy and positive
angular momentum leaves the pair production region and goes towards
the event horizon, much energy and momentum are lost and ultimately
the particle has negative energy and angular momentum \cite{Li2}.
Thus if the particle either with negative or positive energy leaves
the pair production region and gets a chance to reach the event
horizon, the result is the extraction of energy from the black hole
transmitted to the accretion disk. This shows that when the
magnetosphere is filled isothermal plasma admitting the properties
of Veselago as well as usual medium, our results indicate that
energy extraction is possible.

%\newpage
\vspace{0.5cm}

{\bf Acknowledgment}

\vspace{0.5cm}

We appreciate the Higher Education Commission Islamabad, Pakistan,
for its financial support during this work through the {\it
Indigenous PhD 5000 Fellowship Program Batch-II}.

\vspace{0.5cm}

\renewcommand{\theequation}{A\arabic{equation}}
\setcounter{equation}{0}
\section*{Appendix A}
This Appendix includes the details to reach at the perturbed form of
the GRMHD equations (\ref{a1})-(\ref{e1}). The component form of
these equations is also given.

When we introduce the perturbations from Eq.(\ref{pv}), the
linearized GRMHD Eqs.(\ref{a1})-(\ref{e1}) become
\begin{eqnarray}\label{a2}
&&\left\{(\frac{\partial}{\partial t}-\beta.\nabla)(\delta
\textbf{B})\right\}=\nabla\times(\textbf{v}\times\textbf{B})
+\nabla\times(\textbf{V}\times\delta\textbf{B})-(\delta\textbf{B}.\nabla)\beta,\\
\label{b2}
&&\nabla.(\delta \textbf{B})=0,\\
\label{c2} &&\left\{\frac{\partial}{\partial
t}+(\textbf{V}-\beta).\nabla\right\}(\delta\rho+\delta
p)+(\rho+p)\gamma^2\textbf{V}.\left\{\frac{\partial}{\partial
t}+(\textbf{V}-\beta).\nabla\right\}\textbf{v}\nonumber\\
&&+(\rho+p)(\nabla.\textbf{v})+(\delta\rho+\delta
p)(\nabla.\textbf{V})+(\delta\rho+\delta
p)\gamma^2\textbf{V}.(\textbf{V}.\nabla)\textbf{V}
\nonumber\\
&&=-2(\rho+p)\gamma^2(\textbf{V}.\textbf{v})(\textbf{V}.\nabla)\ln
\gamma-(\rho+p)\gamma^2(\textbf{V}.\nabla\textbf{V}).\textbf{v}\nonumber
\end{eqnarray}
\begin{eqnarray}
&&+(\rho+p)\textbf{v}.\nabla \ln u,\\
\label{d2}
&&\left[\left\{(\rho+p)\gamma^2+\frac{\textbf{B}^2}{4\pi}\right\}\delta_{ij}
+(\rho+p)\gamma^4 V_iV_j-\frac{1}{4\pi}B_i
B_j\right]\left(\frac{\partial}{\partial t}-\beta.\nabla
\right)v^j\nonumber\\
&&+(\rho+p)\gamma^2v_{i,j}V^j+(\rho+p)\gamma^4V_iv_{j,k}V^jV^k+\frac{1}{4\pi}
\left[\textbf{B}\times\left\{\textbf{V}\times\frac{d(\delta\textbf{B)}}
{d\tau}\right\}\right]_i\nonumber\\
&&-\frac{1}{4\pi}\left\{(\delta B_i)_{,j}-(\delta
B_j)_{,i}\right\}B^j=-(\delta p)_{,i}+\gamma^2[(\delta\rho+\delta
p)V^j\nonumber\\
&&+2(\rho+p)\gamma^2(\textbf{V}.\textbf{v})V^j+(\rho+p)v^j]\beta_{j,i}
+\frac{1}{4\pi}(B_{i,j}-B_{j,i})\delta
B^j\nonumber\\
&&-(\rho+p)\gamma^4(v_iV^j+v_jV^i)V_{k,j}V^k\nonumber\\
&&-\gamma^2\{(\delta \rho+\delta
p)V^j+2(\rho+p)\gamma^2(\textbf{V}.\textbf{v})V^j+(\rho+p)
v^j\}V_{i,j}\nonumber\\
&&-\gamma^4V_i\{(\delta\rho+\delta
p)V^j+4(\rho+p)\gamma^2(\textbf{V}.\textbf{v})V^j
+(\rho+p)v^j\}V_{j,k}V^k,\\
\label{e2} &&\gamma^2\left\{\frac{\partial}{\partial
t}+(\textbf{V}-\beta).\nabla\right\}(\delta\rho+\delta
p)+2(\rho+p)\gamma^4\textbf{V}.\left\{\frac{\partial}{\partial
t}+(\textbf{V}-\beta).\nabla\right\}\textbf{v}\nonumber\\
&&-(\rho+p)\gamma^2\textbf{v}.\nabla \ln
u+(\rho+p)\gamma^4\textbf{V}.(\textbf{v}.\nabla)\textbf{V}
+6(\rho+p)\gamma^6(\textbf{V}.\textbf{v})\textbf{V}.(\textbf{V}.\nabla)\textbf{V}
\nonumber\\
&&+2(\delta\rho+\delta
p)\gamma^4\textbf{V}.(\textbf{V}.\nabla)\textbf{V}
+2(\rho+p)\gamma^4\textbf{v}.(\textbf{V}.\nabla)\textbf{V}
\nonumber\\
&&-2(\rho+p)\gamma^4(\textbf{V}.\textbf{v})\textbf{V}.\nabla \ln
u+2(\rho+p)\gamma^4(\textbf{V}.\textbf{v})
(\nabla.\textbf{V})-\frac{\partial}{\partial
t}(\delta{p})\nonumber\\
&&+(\delta\rho+\delta
p)\gamma^2(\nabla.\textbf{V})+(\rho+p)\gamma^2(\nabla.\textbf{v})
-\gamma^2(\beta.\nabla)(\delta\rho+\delta
p)\nonumber\\
&&+2(\rho+p)\gamma^4(\textbf{V}.\textbf{v})(\beta.\nabla \ln
u)-6(\rho+p)\gamma^6(\textbf{V}.\textbf{v})\textbf{V}
.(\beta.\nabla)\textbf{V}\nonumber\\
&&-2(\rho+p)\gamma^4\textbf{v}.(\beta.\nabla)\textbf{V}-2(\delta\rho+\delta
p)\gamma^4\textbf{V}.(\beta.\nabla)\textbf{V}
\nonumber\\
&&-(\delta\rho+\delta
p)\gamma^2\textbf{V}.(\textbf{V}.\nabla)\beta-(\rho+p)\gamma^2\textbf{V}.
(\textbf{v}.\nabla)\beta\nonumber\\
&&-2(\rho+p)\gamma^4(\textbf{V}.\textbf{v})
\textbf{V}.(\textbf{V}.\nabla)\beta+\frac{1}{4\pi}\left[(\textbf{v}\times\textbf{B}).(\nabla\times
\textbf{B})\right.\nonumber\\
&&\left.+(\textbf{V}\times\delta\textbf{B}).(\nabla
\times\textbf{B})+(\textbf{V}\times\textbf{B})
.(\nabla\times\delta\textbf{B})\right.\nonumber\\
&&\left.+(\textbf{V}
\times\textbf{B}).\left\{\frac{d\textbf{v}}{d\tau}\times\textbf{B}+\textbf{V}
\times \frac{d \delta \textbf{B}}{d\tau}\right\}\right]=0.
\end{eqnarray}
The component form of these equations are
\begin{eqnarray}\label{a3}
&&\frac{db_x}{d\tau}+Vb_{x,x}+ub_{x,z}=-u'b_x
+(V-\beta)'b_z+v_{x,z}-\lambda v_{z,z}-\lambda'v_z,\\
\label{b3} &&\frac{db_z}{d\tau}+Vb_{z,x}+ub_{z,z}=\lambda
v_{z,x}-v_{x,x},\\\label{c3} &&b_{x,x}+b_{z,z}=0,
\end{eqnarray}
\begin{eqnarray}
\label{d3}
&&\rho\frac{d\tilde{\rho}}{d\tau}+p\frac{d\tilde{p}}{d\tau}+\rho
V\tilde{\rho}_{,x}+pV\tilde{p}_{,x}+\rho
u\tilde{\rho}_{,z}+pu\tilde{p}_{,z}
-(\tilde{\rho}-\tilde{p})\{p'u+pu'\nonumber\\
&&+pu\gamma^2(VV'+uu')\}+(\rho+p)\gamma^2\left(V\frac{dv_x}{d\tau}
+u\frac{dv_z}{d\tau}\right)+(\rho+p)(1+\gamma^2V^2)v_{x,x}\nonumber\\
&&+(\rho+p)(1+\gamma^2u^2)v_{z,z}+(\rho+p)uV\gamma^2(v_{x,z}+v_{z,x})\nonumber\\
&&=-(\rho+p)\gamma^2u[(1+2\gamma^2V^2)V'+2\gamma^2uVu']v_x\nonumber\\
&&+(\rho+p)[(1-2\gamma^2u^2)(1+\gamma^2u^2)\frac{u'}{u}-2\gamma^4u^2VV']v_z,\\
\label{e3}
&&\left\{(\rho+p)\gamma^2(1+\gamma^2V^2)+\frac{B^2}{4\pi}\right\}\frac{dv_x}{d\tau}
+\left\{(\rho+p)\gamma^4uV-\frac{\lambda B^2}{4\pi}\right\}\frac{dv_z}{d\tau}\nonumber\\
&&+\left\{(\rho+p)\gamma^2(1+\gamma^2V^2)-\frac{B^2}{4\pi}\right\}(Vv_{x,x}+uv_{x,z})
-\frac{B^2}{2\pi}uVb_{z,z}\nonumber\\
&&+\left\{(\rho+p)\gamma^4uV+\frac{\lambda
B^2}{4\pi}\right\}(Vv_{z,x}+uv_{z,z})+\frac{B^2}{4\pi}\{(1-V^2)b_{z,x}\nonumber\\
&&-(1-u^2)b_{x,z}\}=-\frac{B^2}{4\pi}uu'b_x+\frac{B^2}{4\pi}\{\lambda'+u(V-\beta)'\}b_z
-p\tilde{p}_{,x}\nonumber\\
&&-(\rho\tilde{\rho}+p\tilde{p})\gamma^2 u\{(1+\gamma^2
V^2)V'+\gamma^2uVu'\}-(\rho+p)\gamma^4
u\{(1+4\gamma^2V^2)uu'\nonumber\\
&&+4(1+\gamma^2V^2)VV'\}v_x-[(\rho+p)\gamma^2[\{(1+2\gamma^2
u^2)(1+2\gamma^2 V^2)-\gamma^2
V^2\}V'\nonumber\\
&&+2\gamma^2(1+2\gamma^2
u^2)uVu']+\frac{B^2}{4\pi}u\lambda']v_z,\\\label{f3}
&&\left\{(\rho+p)\gamma^2(1+\gamma^2u^2)
+\frac{\lambda^2B^2}{4\pi}\right\}\frac{dv_z}{d\tau}
+\left\{(\rho+p)\gamma^4uV
-\frac{\lambda B^2}{4\pi}\right\}\frac{dv_x}{d\tau}\nonumber\\
&&+\left\{(\rho+p)\gamma^2(1+\gamma^2u^2)
-\frac{\lambda^2B^2}{4\pi}\right\}(Vv_{z,x}
+uv_{z,z})+\frac{\lambda B^2}{2\pi}uVb_{z,z}\nonumber\\
&&+\left\{(\rho+p)\gamma^4uV +\frac{\lambda
B^2}{4\pi}\right\}(Vv_{x,x}+uv_{x,z})-\frac{\lambda
B^2}{4\pi}\{(1-V^2)b_{z,x}\nonumber\\
&&-(1-u^2)b_{x,z}\}=-\frac{B^2}{4\pi}(\lambda'-\lambda uu')b_x
-\frac{\lambda B^2}{4\pi}u(V-\beta)'b_z\nonumber\\
&&-\rho\tilde{\rho}\gamma^2\{uu'(1+\gamma^2u^2)+\gamma^2u^2VV'-V\beta'\}
-[p\tilde{p}_{,z}+p'\tilde{p}\nonumber\\
&&+p\tilde{p}\gamma^2\{uu'(1+\gamma^2u^2)+\gamma^2u^2VV'-V\beta'\}]
-(\rho+p)\gamma^2[\gamma^2u^2(1+4\gamma^2V^2)V'\nonumber\\
&&-(1+2\gamma^2 V^2)\beta'+2\gamma^2uV(1+2\gamma^2
u^2)u']v_x-[(\rho+p)\gamma^2\{-2\gamma^2uV\beta'\nonumber\\
&&+(1+\gamma^2u^2)(1+4\gamma^2u^2)u'+2\gamma^2(1+2\gamma^2u^2)uVV'
\}-\frac{\lambda B^2}{4\pi}u\lambda']v_z,\\
\label{g3} &&\gamma^2\rho\frac{\partial \tilde{\rho}}{\partial
t}+p(\gamma^2-1)\frac{\partial \tilde{p}}{\partial
t}+\tilde{\rho}\gamma^2\{\rho'u+\rho u'+2\rho u
\gamma^2(VV'+uu')\nonumber
\end{eqnarray}
\begin{eqnarray}
&&-\rho uV\beta'\}+\tilde{p}\gamma^2\{up'+u'p+2pu\gamma^2(VV'+uu')
-p\gamma^2uV\beta'\}\nonumber\\
&&+\gamma^2\rho\tilde{\rho}_{,x}(V-\beta)
+\gamma^2u\rho\tilde{\rho}_{,z}+\gamma^2p\tilde{p}_{,x}(V-\beta)
+\gamma^2up\tilde{p}_{,z}\nonumber\\
&&+\frac{\partial v_x}{\partial
t}\{2(\rho+p)\gamma^4V-\frac{B^2}{4\pi}(u\lambda-V)\}+\frac{\partial
v_z}{\partial t}\{2(\rho+p)\gamma^4u+\frac{\lambda
B^2}{4\pi}(u\lambda-V)\}\nonumber\\
&&+v_{x,x}[(\rho+p)\gamma^2\{1+2\gamma^2V(V-\beta)\}
-\frac{B^2}{4\pi}(V-\beta)(u\lambda-V)]\nonumber\\
&&+v_{x,z}u[2(\rho+p)\gamma^4V
+\frac{B^2}{4\pi}(u\lambda-V)]+v_{z,x}(V-\beta)[2(\rho+p)\gamma^4u
\nonumber\\
&&+\frac{B^2\lambda}{4\pi}(u\lambda-V)]+v_{z,z}\{(\rho+p)(1+2\gamma^2u^2)
-\frac{B^2\lambda}{4\pi}u(u\lambda-V)\}\nonumber\\
&&+\frac{B^2}{4\pi}(u\lambda-V) \{(1-u^2)b_{x,z}-(1-V^2)b_{z,x}
+2uVb_{z,z})\}\nonumber\\
&&+\frac{B^2}{4\pi}ub_x\{\lambda'-(u\lambda-V)u'\}
+\frac{B^2}{4\pi}b_z\{-\lambda'V+u(u\lambda-V)(V-\beta)'\}\nonumber\\
&&+v_x[(\rho+p)\gamma^2u\{2\gamma^2V'+6\gamma^4V(VV'+uu')-\beta'(1+2\gamma^2V^2)\}
-\frac{B^2\lambda'}{4\pi}]\nonumber\\
&&+v_z[\frac{B^2\lambda'}{4\pi}\{\lambda-u(u\lambda-V)\}
+(\rho+p)\gamma^2\{-\frac{u'}{u}+2\gamma^2uu'+6\gamma^4u^2(VV'+uu')\nonumber\\
&&+\gamma^2(VV'+uu')-V\beta'(1+\gamma^2u^2)\}]=0.
\end{eqnarray}
We have used the conservation law of rest-mass for three-dimensional
hypersurface, i.e., given by Eq.(\ref{clm}) to simplify
Eq.(\ref{d3}). The same equation will be used to simplify the
Fourier analyzed form of Eqs (4.11)-(4.17).


\vspace{0.1cm}

\begin{thebibliography}{99}

\bibitem{ADM}Arnowitt, R., Deser, S. and Misner, C.W.:
\textit{Gravitation: An Introduction to Current Research} ed.
Witten, L. (Wiley, New York, 1962).

\bibitem{TM1}Thorne, K.S. and Macdonald, D.A.: Mon. Not. R. Astron. Soc.
\textbf{198}(1982)339.

\bibitem{TM2}Thorne, K.S. and Macdonald, D.A.: Mon. Not. R. Astron. Soc.
\textbf{198}(1982)345.

\bibitem{TPM}Thorne, K.S., Price, R.H. and Macdonald, D.A.: \textit{Black Holes:
The Membrane Paradigm} (Yale University Press, New Haven, 1986).

\bibitem{HT}Holcomb, K.A. and Tajima, T.: Phys. Rev. \textbf{D40}(1989)3809.

\bibitem{Ho}Holcomb, K.A.: Astrophys. J. \textbf{362}(1990)381.

\bibitem{De}Dettmann, C.P., Frankel, N.E. and Kowalenko, V.:  Phys. Rev.
\textbf{D48}(1993)5655.

\bibitem{Kom} Komissarov, S.S.: Mon. Not. R. Astron. Soc. \textbf{350}(2004)427.

\bibitem{BZ} Blandford, R.D. and Znajek, R.L.: Mon. Not. R. Astron. Soc.
\textbf{179}(1977)433.

\bibitem{P} Penrose, R.: Rev. Nouvo Cim. \textbf{1}(1969)252.

\bibitem{G} Goldreich, P. and Julian, W.H.: Astrophys. J. \textbf{157}(1969)869.

\bibitem{Can} Canuto, V. and Chiuderi, C.: Phys. Rev. \textbf{D1}(1970)2219.

\bibitem{Pre} Press, W.H. and Teukolsky, S.A.: Nature \textbf{238}(1972)211.

\bibitem{Mu} M\"{u}ller, A.: D.Sc. Dissertation (Ruperto-Carola University of
Heidelberg, 2004).

\bibitem{Pu} Semenov, V., Dyadechkin, S. and Punsly, B.: Science \textbf{305}(2004)978;\\
Chicone, C., Mashhoon, B. and Punsly, B.: Int. J. Mod. Phys. \textbf{D13}(2003)945;\\
Punsly, B.: Astrophys. J. \textbf{583}(2003)842;\\
Chicone, C., Mashhoon, B. and Punsly, B.: Phys. Lett. \textbf{A343}(2005)1;\\
Punsly, B.: Mon. Not. R. Astron. Soc. \textbf{366}(2006)29.

\bibitem{MK} Musil, T. and Karas, V.: Publ. Astron. Soc. Jap.
\textbf{54}(2002)641.

\bibitem{Ko}Koide, S., Shibata, K., Kudoh, T. and Meier, D.L.:
Science \textbf{295}(2002)1688.

\bibitem{Z1}Zhang, X.-H.: Phys. Rev. \textbf{D39}(1989)2933.

\bibitem{Z2}Zhang, X.-H.: Phys. Rev. \textbf{D40}(1989)3858.

\bibitem{BH1}Buzzi, V., Hines, K.C. and Treumann, R.A.: Phys. Rev.
\textbf{D51}(1995)6663.

\bibitem{BH2}Buzzi, V., Hines, K.C. and Treumann, R.A.: Phys. Rev.
\textbf{D51}(1995)6677.

\bibitem{M} Mackay, T.G., Lakhtakia, A. and Setiawan, S.: New J. Phys.
\textbf{7}(2005)171.

\bibitem{V}Veselago, V.G.: Sov. Phys. Usp.
\textbf{10}(1968)509.

\bibitem{SSS} Shelby, R.A., Smith, D.R. and Schultz, S.: Science \textbf{292}(2001)77.

\bibitem{many} Leonhardt, U., Valanju, P.M., Wasler, R.M. and Valanju, A.P.: Phys. Rev. Lett.
\textbf{88}(2002)187401-1; IEEE Journal of Selected Topics in
Quantum Electronics, \textbf{9}(2003)102;\\
Dolling, G., Enkrich, C., Wegener, M., Soukoulis, C.M. and Linden
S.: Science \textbf{312}(2006)892.

\bibitem{WM} Woodley, J.F. and Mojahedi, M.: \emph{Negative Group Velocity in
Left-Handed Materials}, Antennas and Propagation Society
International Symposium and USNC/CNC/URSI National Radio Science
Meeting, Columbus, Ohio, USA, June 22-27, 2003, Vol. 4, p 643.

\bibitem{S1} Sharif, M. and Sheikh, U: Gen. Relat.
Gravit. \textbf{39}(2007)1437.

\bibitem{S2} Sharif, M. and Sheikh, U: \textit{Effects of Schwarzschild Black Hole
Horizon on Isothermal Plasma Wave Dispersion}, Gen. Relat. Gravit.
(2007, to appear) gr-qc 0708.2690.

\bibitem{S6}Sharif M. and Sheikh, U.: \textit{Cold Plasma Gravitomagnetic
Waves in Kerr Planar Analogue}, submitted for publication.

\bibitem{Ja} Jackson, J.D.: \textit{Classical Electrodynamics}
(Wiley, New Jersey, 1999).

\bibitem{AL} Abramowicz, M.A. and Lacosta, J.P.: Acta Astron. \textbf{30}(1980)1.

\bibitem{Li2} Li, L.-X.: Astrophys. J. \textbf{ L17}(2000)540.

\end{thebibliography}
\end{document}